\documentclass[a4paper,12pt]{article}
\pdfoutput=1         
\usepackage{jcappub}        
\usepackage[T1]{fontenc} 
\usepackage{subfigure}
\definecolor{ultramarine}{rgb}{0.07, 0.04, 0.56}
\definecolor{cadmiumgreen}{rgb}{0.0, 0.42, 0.24}
\definecolor{indigo(dye)}{rgb}{0.0, 0.25, 0.42}
\usepackage[linktocpage=true]{hyperref}
\hypersetup{
colorlinks=true,
citecolor=ultramarine,
linkcolor=indigo(dye),
urlcolor=indigo(dye),
pdfauthor={},
pdftitle={},
pdfsubject={}
}

\title{Instability of De Sitter Spacetime induced by Quantum Conformal Anomaly}

\author[a]{Hiroki Matsui,}
\affiliation[a]{Department of Physics, Tohoku University, Sendai, Miyagi 980-8578, Japan}
\emailAdd{hiroki.matsui.c6@tohoku.ac.jp}

\abstract{The instability of (quasi) de Sitter spacetime from 
quantum gravitational effects has been discussed in many works.
Especially, the gravitational backreaction from quantum energy momentum tensor 
is crucial for understanding the low-energy description of 
quantum gravity and sometimes destabilize the spacetime. 
In this paper we discuss the (quasi) de Sitter instability 
from gravitational backreaction involving quantum conformal anomaly.
The conformal or trace anomaly corresponds to the 
quantum gravitational contributions of the massless conformal fields
and affects the spacetime homogeneously.
First, we derive the conformal anomaly using the adiabatic (WKB) approximation
and discuss the renormalization of the quantum energy momentum tensor.
Then, we consider the dynamics of the Hubble parameter 
based on the semiclassical Einstein's equations including 
the cosmological constant, the conformal anomaly and the higher-derivative terms.
We have clearly shown that the classical de Sitter attractor $H_{\mathrm{C}} \simeq \sqrt{{\Lambda}/{3}}$
are generally unstable from the viewpoint of the semiclassical gravity
and the inflation is destabilized except for the specific conditions.
Unless the fine-tuning of the conformal anomaly and 
the higher derivative terms, the inflation 
finally becomes the Planckian inflation with the Hubble scale $H \approx M_{\rm P}\equiv \sqrt{1/8\pi G_{N}}$ or terminates $H(t) \rightarrow 0$. 
The latter case suggests that the cosmic inflation 
could not last long and the eternal inflation 
scenarios are strongly constrained.}

\begin{document}
\maketitle
\flushbottom
\newcommand{\Slash}[1]{{\ooalign{\hfil/\hfil\crcr$#1$}}}

\section{Introduction}
\label{sec:intro}
The modern gravitational physics is mainly based on classical Einstein's theory. 
However, sometimes in early Universe and black hole physics, 
we face a necessity to properly handle quantum gravitational  phenomena. 
In principle, the quantum phenomena involving gravity should be discussed in 
the framework of quantum gravity (QG) theory 
where metric is also quantized together with matter fields. 
However, there has been a notorious problem to devise a consistent theory of QG 
and we have no complete solution~\cite{Smolin:2003rk,DeWitt:2007mi,Giddings:2011dr}.
As the most important and well-known approximation of QG, 
the semiclassical gravity where only matter fields are quantized,
but metric is treated as a classical background, 
provides a satisfactory description~\cite{Birrell:1982ix} and 
there are many successful examples.
Especially, including backreaction effects of the quantum fluctuations onto the spacetime
is crucial for understanding the quantum nature of the gravity 
and the evaporation of the back hole~\cite{Hawking:1974sw}.

The effective action of gravity is defined through the path integral
over the set of all matter fields $\phi$ including ghosts and the gravity field $g_{\mu\nu}$,
\begin{align}
e^{i\Gamma_{\rm eff} \left[g_{\mu\nu}\right]}\,=\int { \mathcal{D}\phi \,e^{iS\left[\phi,\,g_{\mu\nu}\right]} } ,
\end{align}
where the classical action $S\left[\phi,\,g_{\mu\nu}\right]$~\cite{Buchbinder:1992rb}
is formally defined by
$S\left[\phi,\,g_{\mu\nu}\right]=S_{\rm gravity}\left[g_{\mu\nu}\right]
+S_{\rm matter}\left[\phi,\,g_{\mu\nu}\right]$
including all matter fields $\phi$, their couplings and the metric.
The semiclassical gravity has no unitary problem about 
the gravitational $S$-matrix~\cite{Salles:2014rua} since the gravity is not quantized.
However, the renormalization of the effective action requires 
the Einstein-Hilbert action including a cosmological constant,
\begin{align}
S_{EH} =  -\frac { 1 }{ 16\pi { G }_{ N } } \int { { d }^{ 4 }x\sqrt { -g } \left( R+2\Lambda  \right)  },
\label{EH}
\end{align}
and the high-order gravitational action
\begin{align}
S_{HG}=  -\int { { d }^{ 4 }x\sqrt { -g } \left( { a }_{ 1 }F+{ a }_{ 2 }E+{ a }_{ 3 }\Box R  \right)  },
\label{HD}
\end{align}
where $a_{1},a_{2},a_{3}$ are high-order derivative couplings.
Note that $S_{HG}$ should be required to have a renormalizable theory in curved spacetime.
$E$ is the Gauss-Bonnet invariant term and $F$ is the square of the Weyl tensor
defined by the Riemann curvature tensor $R_{\mu\nu\kappa\lambda}$, 
the dual Riemann curvature tensor
$\mbox{}^{*}R_{\mu\nu\kappa\lambda}=\varepsilon_{\mu\nu\alpha\beta}
R^{\alpha\beta}_{\phantom{\alpha\beta}\kappa\lambda}/2$
and the Weyl tensor $C_{\mu\nu\kappa\lambda}$,
\begin{subequations}
\label{TraceAnomaly2}
\begin{align}
E &\equiv \mbox{}^{*}R_{\mu\nu\kappa\lambda}
\mbox{}^{*}R^{\mu\nu\kappa\lambda} = R_{\mu\nu\kappa\lambda}
R^{\mu\nu\kappa\lambda}- 4 R_{\mu\nu}
R^{\mu\nu} + R^{2} \label{TraceAnomaly2a}, \\
F &\equiv C_{\mu\nu\kappa\lambda} C^{\mu\nu\kappa\lambda} =
R_{\mu\nu\kappa\lambda} R^{\mu\nu\kappa\lambda}- 2 R_{\mu\nu}
R^{\mu\nu} + \frac{1}{3} R^{2} \label{TraceAnomaly2b},
\end{align}
\end{subequations}
The principle of least action with respect to the total action
yields general Einstein's equations
properly dealing with the quantum effects below the Planck scale,
\begin{align}\label{eq:semiclassical}
\frac { 1 }{ 8\pi { G }_{ N } } { G }_{ \mu\nu  }+{ \rho  }_{ \Lambda  } 
 { g }_{ \mu\nu }+{ a }_{ 1 }{ H }_{ \mu\nu }^{ \left( 1 \right)  }
+{ a }_{ 2 }{ H }_{ \mu\nu  }^{ \left( 2 \right)  }+{ a }_{ 3 }{ H }_{ \mu\nu  }^{ \left( 3 \right)  }
=\left< { T }_{ \mu\nu } \right>,
\end{align}
where ${ G }_{ \mu \nu  }={ R }_{ \mu \nu  }-\frac { 1 }{ 2 }R{ g }_{ \mu \nu  }$ is the Einstein tensor and 
${ \rho  }_{ \Lambda  } \equiv \Lambda /{8\pi { G }_{ N }}$ defined by the cosmological constant $\Lambda$, 
$R_{\mu\nu}$ and $R$ are the Ricci tensor and scalar, ${ H }_{ \mu \nu  }^{ \left( 1 \right)  }$, ${ H }_{ \mu \nu  }^{ \left( 2 \right)  }$
or ${ H }_{ \mu \nu  }^{ \left( 3 \right)  }$ are covariantly conserved tensors and 
$\left< { T }_{ \mu \nu  } \right>$ is the vacuum expectation values of 
the energy momentum tensor ${ T }_{ \mu \nu  }$.
Note that the quantum energy momentum tensor 
deforms the background spacetime: $\left< { T }_{ \mu \nu  } \right>
\Longrightarrow { g }_{ \mu \nu  }$.
The vacuum expectation values of
the energy momentum tensor $\left< { T }_{ \mu \nu  } \right>$
include quantum radiative corrections and also gravitational particle production effects in curved spacetime~\cite{Birrell:1982ix}.
The quantum energy momentum tensor $\left< { T }_{ \mu \nu  } \right>$ strongly affects the background spacetime
and the quantum backreaction on the metric ${ g }_{ \mu \nu  }$ comes from these twofold effects.

The gravitational backreaction from the 
quantum energy momentum tensor $\left< { T }_{ \mu \nu  } \right>$ is 
crucial for understanding the low-energy description of QG
and sometimes destabilize the background spacetime.
For instance, the instability of de Sitter spacetime has been discussed
based on the quantum particle creations of minimally coupled massless scalar or graviton~\cite{Ford:1984hs,Mottola:1984ar,Mottola:1985qt,Antoniadis:1985pj,Antoniadis:1986sb,Higuchi:1986ww,Polarski:1990tr,Tsamis:1992sx,Tsamis:1994ca,
Tsamis:1992xa,Tsamis:1996qq,Tsamis:1996qk,Mukhanov:1996ak,Abramo:1997hu,Goheer:2002vf,Brandenberger:2002sk,Kachru:2003aw,Finelli:2004bm,Janssen:2007ht,Janssen:2008dw,Janssen:2008dp,Polyakov:2009nq,Shukla:2016bnu,Anderson:2013ila,Anderson:2013zia}
and the thermal feature of cosmological horizon~\cite{Gibbons:1977mu,Dyson:2002pf,Padmanabhan:2002ji,Padmanabhan:2003gd,ArkaniHamed:2007ky,Polyakov:2007mm,Markkanen:2016aes,Markkanen:2016jhg,Markkanen:2016vrp,Markkanen:2017abw,Dvali:2017eba}
The continuous particle production or purely thermodynamic description of de Sitter spacetime
imply that de Sitter spacetime might not  be stable.
However, the instability of the de Sitter spacetime is inconsistent with the naive consideration 
and has been still under debate~\cite{Horowitz:1978fq,Simon:1991bm,
Parker:1993dk,Anderson:2002fk,Ford:2005qz,
Frob:2012ui,Frob:2013ht}.

In the present paper we discuss the de Sitter instability from quantum backreaction 
involving the conformal anomaly~\cite{Capper:1974ic,Deser:1976yx,Duff:1977ay,Christensen:1978gi}.
The conformal or trace anomaly corresponds to the quantum contributions of the massless conformal particles and 
one of the most interesting phenomena of the quantum field theory (QFT) 
in curved spacetime~\cite{Duff:1993wm}.
Famously, the conformal field theory (CFT) such as the electromagnetism or 
the conformally massless scalar field theory has the vanishing trace of the stress
tensor classically: $T_{\phantom{\mu}\mu}^{\mu}=0$. 
But, the vacuum expectation values of the stress
tensor $\left\langle T_{\phantom{\mu}\mu}^{\mu}\right\rangle$ 
becomes non-zero and 
the classical property disappears in the QFT.
The general form of conformal anomaly for four dimensions
in curved spacetime is given by \cite{Birrell:1982ix, Capper:1974ic,Deser:1976yx,Duff:1977ay,Christensen:1978gi}:
\begin{align}\label{eq:Traceanomaly}
\left\langle
T_{\phantom{\mu}\mu}^{\mu}\right\rangle_{\rm anomaly} = b F + b' E + c\, \Box R \,,
\end{align}
where $b$, $b'$ and $c$ are dimensionless parameters originating from 
massless conformally covariant fields with different spin values.
The derivation of the conformal anomaly using the dimensional regularization~\cite{Buchbinder:1992rb}, 
the point-splitting~\cite{Christensen:1978yd} and $\zeta$-regularization~\cite{Dowker:1976zf}
lead to the definite results except for the coefficient of the $\Box R$ term.
The ambiguity of conformal anomaly comes from
the parameter $c$ which is regularization-scheme dependent and 
also gauge dependent. 
The geometric properties of the conformal anomaly undoubtedly add the higher-derivatives 
corrections to the ordinary Einstein equations
although the regularization ambiguity reduces the self-consistent description about  
the quantum backreaction.

First, in this paper we consider adiabatic (WKB) approximation for the conformally massless fields
and derive the conformal anomaly in this method.
We clearly show that the derivation of adiabatic (WKB) approximation
reduces the ambiguity of the conformal anomaly and it becomes more physical than any other regularization 
formalism by considering the adiabatic vacuum $\left| 0 \right>_{\rm A} $.
After review and several discussions of the conformal anomaly, 
we investigate how the gravitational backreaction involving 
the conformal anomaly detribalizes the (quasi) de Sitter spacetime. 
The conformal anomaly has a nontrivial impact on the spacetime 
and has been widely discussed in many context.
For instance, the conformal anomaly can relax a big bang or crunch singularity of the 
Friedmann-Lemaitre-Robertson-Walker (FLRW) spacetime~\cite{Nojiri:2004ip,Awad:2015syb}.
It has been argued that the conformal anomaly potentially provide a dynamical solution of  
the cosmological constant problem~\cite{Tomboulis:1988gw,Antoniadis:1991fa,Antoniadis:1998fi,
Schutzhold:2002pr,Antoniadis:2006wq,Bilic:2007gr,Koksma:2008jn}.
As the another application, the conformal anomaly can provide the satisfactory scenario of the inflation. 
Originally, Starobinsky~\cite{Starobinsky:1980te} introduced that the conformal anomaly can be a source of the cosmic de Sitter stage.
Subsequently, this theory has been established as so-called anomaly induced inflation~\cite{Hawking:2000bb,Shapiro:2001rh,
Pelinson:2002ef,Netto:2015cba} from a more modern point of view.
Based on these works we investigate the stability of the de Sitter spacetime 
with the conformal anomaly, the higher-derivative terms and the cosmological constant in more detail.
We clearly showed that the cosmological constant $\Lambda$
does not determine the ultimate dynamics
of the spacetime except for the initial behavior.
We focus on the CFT, but the generalization is not difficult.
Our results suggest 
that the classic de Sitter spacetime dominated by $\Lambda$ is not stable 
from the viewpoint of the semiclassical gravity 
and the cosmic inflation is generally destabilized except for the specific conditions. 
Unless the fine-tuning of the conformal anomaly and the
higher derivative terms, the initial inflation finally becomes
the Planckian inflation with the Hubble scale $H \approx M_{\rm P}\equiv \sqrt{1/8\pi G_{N}}$ 
or terminates $H(t) \rightarrow 0$. This fact suggests that cosmic inflation should not last long and 
some scenarios of eternal inflation must be revisited.

The present paper is organized as follows. 
In Section~\ref{sec:backreaction}, we review some technical difficulties 
of the renormalization 
of the quantum energy momentum tensor and 
discuss the adiabatic (WKB) approximation for the conformally 
massless scalar field. We consider the regularization-scheme 
dependence of the conformal anomaly
and clearly show the ambiguity can be reduced in the adiabatic approximation.
In Section~\ref{sec:instability}, we consider the stability of the de Sitter spacetime 
with the conformal anomaly and the higher-derivative terms.
We clearly show that the classic de Sitter spacetime is generally unstable and the inflation
rolls down to the Planckian stage or terminates. 
We clearly showed that the cosmological constant 
does not determine the ultimate dynamics
of the spacetime except for the initial behavior.
In Section~\ref{sec:cosmic_inflation}, we discuss the cosmological application of the de Sitter instability
considered in previous Section~\ref{sec:instability}  
and especially focus on eternal new inflation or chaotic inflation.
Finally, in Section~\ref{sec:conclusion} we draw the conclusion of our work.

\section{Gravitational backreaction from conformally scalar field }
\label{sec:backreaction}
Let us consider quantum gravitational effects in the early Universe. 
In order to consider the gravitational backreaction involving quantum conformal anomaly quantitatively,
we give a brief review of the renormalization 
issues of the quantum energy momentum tensor
and discuss the adiabatic (WKB) approximation for conformally massless scalar field. 
Here, we derive the conformal anomaly using the adiabatic approximation and 
clearly show that the conformal anomaly obtained by this formulation 
is more simple than any other regularization.
Although the conformal anomaly includes a regularization dependent parameter $c$,
the ambiguity of the conformal anomaly for the coefficient of the $\Box R$ term
can be removed by considering the adiabatic vacuum $\left| 0 \right>_{\rm A} $ 
and its formulation.

Let us assume the matter action for the conformally coupled scalar field $\phi$ 
in curved spacetime,
\begin{align}
S=\int { { d }^{ 4 }x\sqrt { -g } \left( -\frac { 1 }{ 2 } { g }^{ \mu \nu  }
{ \partial   }_{ \mu  }\phi {\partial  }_{ \nu  }\phi -\frac{1}{2}\left(m^{2}+\frac{R}{6} \right)\phi^{2}  \right)  } \label{eq:dddddddfg},
\end{align}
which lead to the Klein-Gordon equation given as 
\begin{align}
\Box \phi -\left(m^{2}+ \frac{R}{6} \right)\phi  =0 \label{eq:dsssssdg},
\end{align}
where
$\Box =g^{\mu\nu}{ \nabla  }_{ \mu  }{  \nabla   }_{ \nu  }$ express generally covariant d'Alembertian operator.
The energy momentum tensor can be given by~\cite{Bunch:1980vc}
\begin{align}
\begin{split}
{ T }_{ \mu \nu  }&=\frac { -2 }{ \sqrt { -g }  } \frac { \delta S  }{ \delta { g }^{ \mu \nu  } }=
\frac{2}{3}{ \partial  }_{ \mu  }\phi { \partial  }_{ \nu  }\phi -\frac{1}{6}{ g }_{ \mu \nu  }{ g }^{ \rho \sigma  }{ { \partial  }_{ \rho  }\phi \partial  }_{ \sigma  }\phi 
-\frac{1}{3} \phi\nabla_{ \mu  }\nabla_{ \nu  }\phi \\ &+\frac{1}{3} { g }_{ \mu \nu  }{ \phi \Box \phi 
-\frac{1}{6} { G }_{ \mu \nu  }{ \phi  }^{ 2 }+\frac { 1 }{ 2 } { m }^{ 2 }{ g }_{ \mu \nu  }{ \phi  }^{ 2 } },
\end{split}
\end{align}
and the trace
\begin{align}
T_{\phantom{\mu}\mu}^{\mu}={ m }^{ 2 }{ \phi  }^{ 2 } ,
\end{align}
which is exactly zero when $m\rightarrow 0$.
Thus, the conformally massless scalar field 
has the vanishing trace of the stress tensor classically.

\subsection{Adiabatic (WKB) approximation for conformally scalar field }
\label{sec:adiabatic}
Let us consider flat Friedmann-Lemaitre-Robertson-Walker
(FLRW) spacetimes where the metric is given by
$g_{\mu\nu}=\mathrm{diag}\left( -1,{ a }^{ 2 }\left( t \right)  ,{ a }^{ 2 }\left( t \right) { r }^{ 2 },
{ a }^{ 2 }\left( t \right) { r }^{ 2 }\sin^{2} { \theta  }  \right)$ and assume that $a(t)$ is the scale 
factor of the Universe and $t$ is the cosmic time.
Next, let us decompose the scalar field  $\phi \left(\eta ,x\right)$ into the classic part and the quantum part as
\begin{align}
\phi \left(\eta ,x\right)=\phi \left(\eta\right)+\delta \phi \left(\eta,x \right) \label{eq:kkkkgedg},
\end{align}
where $\eta$ is the conformal time defined by $d\eta=dt/a$.
The vacuum expectation value (VEV) of the scalar field satisfy
$\phi\left(\eta \right)=\left< 0  \right| { \phi \left(\eta ,x\right) }\left| 0 \right>$ and 
$\left< 0  \right| { \delta\phi \left(\eta ,x\right) }\left| 0 \right>=0$.
The quantum field $\delta \phi \left( \eta ,x \right)$ can be decomposed into each $k$ modes as follows:
\begin{align}
\delta \phi \left( \eta ,x \right) =\int { { d }^{ 3 }k\left( { a }_{ k }\delta { \phi  }_{ k }\left( \eta ,x \right) 
+{ a }_{ k }^{ \dagger  }\delta { \phi  }_{ k }^{ * }\left( \eta ,x \right)  \right)  }  \label{eq:ddfkkfledg},
\end{align}
where we introduce,
\begin{align}
{ \delta \phi  }_{ k }\left( \eta ,x \right) =\frac { { e }^{ ik\cdot x } }{ { \left( 2\pi  \right)  }^{ 3/2 }
\sqrt { C\left( \eta  \right)  }  } \delta { \chi  }\left( \eta,k  \right)  \label{eq:slkdlkgdg},
\end{align}
and $C\left(\eta \right)=a^{2}\left(\eta \right)$.
The quantum mode function $\delta \chi\left( \eta,k  \right)$ satisfies the 
second-order differential equation which is given by,
\begin{align}
{ \delta \chi}''\left( \eta,k  \right)+{ \omega  }_{ k }^{ 2 }\left( \eta  \right)  \delta \chi\left( \eta,k  \right)=0 \label{eq:dlkrlekg},
\end{align}
where prime express the differential with respect to the conformal time $\eta$ and 
${ \omega  }_{ k }^{ 2 }\left( \eta  \right) ={ k }^{ 2 }+C\left( \eta  \right){ m }^{ 2 } $.
The mode function $\delta \chi\left( \eta,k  \right)$ should satisfy the Wronskian condition,
\begin{align}
{ \delta\chi  }\left( \eta,k  \right){ \delta\chi  }^{ * }\left( \eta,k  \right)
-{ \delta\chi  }^{ * }\left( \eta,k  \right){ \delta\chi  }\left( \eta,k  \right)= i\label{eq:dddrrrg}.
\end{align}
which ensures that the canonical commutation relations.
The canonical commutation relations for the field operator $\delta\chi$ are given by,
\begin{align}
\bigl[ { a }_{ k },{ a }_{ k' } \bigr] = \bigl[{ a }_{ k }^{ \dagger  },{ a }_{ k' }^{ \dagger  } \bigr]=0, \quad 
\bigl[{ a }_{ k },{ a }_{ k' }^{ \dagger  } \bigr] =\delta \left( k-k' \right) \label{eq:oeuoegedg}.
\end{align}
In curved spacetime the vacuum state annihilated by all the operators ${ a }_{ k }$
is determined by the choice of the mode functions. 
However, solving Eq.~(\ref{eq:dlkrlekg}) is analytically impossible and we usually adopt a reasonable or continental approximation. 
The adiabatic (WKB) approximation to the mode function ${ \delta\chi  }\left( \eta,k  \right)$ 
is  written by~\cite{Parker:1974qw}:
\begin{align}
{ \delta\chi  }\left( \eta,k  \right)=\frac { 1 }{ \sqrt { 2{ W }_{ k }\left( \eta  \right) C\left( \eta  \right)  }  } 
\left( \alpha_{ k }\cdot e^{-i\int { { W }_{ k }\left( \eta  \right) \, d\eta  }}+
\beta_{ k }\cdot e^{i\int { { W }_{ k }\left( \eta  \right)\, d\eta  }}\right) \label{eq:jgskedg},
\end{align}
where $\alpha_{ k }$ and $\beta_{ k }$ are coefficients satisfying the following condition
\begin{align}
{ { \left| { \alpha  }_{ k } \right|  }^{ 2 } }-{ \left| { \beta  }_{ k } \right|  }^{ 2 }=1.
\end{align}
From Eq.~(\ref{eq:dlkrlekg}) the adiabatic function ${ W }_{ k }\left( \eta  \right)$ must 
satisfy the differential equation~\cite{Bunch:1980vc},
\begin{align}
{ W }_{ k }^{ 2 }={ \omega  }_{ k }^{ 2 }-\left(\frac { 1 }{ 2 } \frac { { W }''_{ k } }{ { W }_{ k } } 
-\frac { 3 }{ 4 } \frac { { \left( { W }'_{ k } \right)  }^{ 2 } }{ { W }_{ k }^{ 2 } }\right) \label{eq:sklsdkldg},
\end{align}
which is analytically impossible to solve. 
But if the background is slowly changing and 
the adiabatic (WKB) conditions (${ \omega  }_{ k }^{ 2 }>0$ and 
$\left| { \omega ' }_{ k }/{ \omega  }_{ k }^{ 2 } \right| \ll 1$) are satisfied, 
we can obtain the adiabatic solution by solving iteratively Eq.~(\ref{eq:sklsdkldg}).
The lowest-order adiabatic solution ${ W }^{( 0 )}_{ k }$ is given by~\cite{Bunch:1980vc}:
\begin{align}
\left({ W }^{( 0 )}_{ k }\right)^{2}={ \omega  }_{ k }^{2}\label{eq:flkldg}.
\end{align}
The first-order adiabatic solution ${ W }^{( 1 )}_{ k }$ is given by 
\begin{align}
\left({ W }^{( 1 )}_{ k }\right)^{2}&={ \omega  }_{ k }^{2}-\frac { 1 }{ 2 } \frac { \left({ W}^{( 0 )}_{ k }\right)'' }{{ W }^{( 0 )}_{ k } } 
+\frac { 3 }{ 4 } \frac {\left({{ W}^{( 0 )}_{ k }}'\right)^{2}  }{ \left({ W}^{( 0 )}_{ k }\right)^{2} }  \\
&={ \omega  }_{ k }^{2}- \frac { 1 }{ 4 }\frac { m^2C'' }{ { \omega  }_{ k }^2 } 
+\frac { 1 }{ 16 }\frac { m^4C'^{2}  }{ { \omega  }_{ k }^{4} }
\label{eq:skdl;lgedg}.
\end{align}
For the high-order adiabatic solution, we obtains the following expression 
\begin{align}
{ W }_{ k }&\simeq{ \omega  }_{ k }
-\frac { { m }^{ 2 }C }{ 8{ \omega  }_{ k }^{ 3 } } \left( D'+{ D }^{ 2 } \right) +\frac { 5{ m }^{ 4 }{ C }^{ 2 }{ D }^{ 2 } }{ 32{ \omega  }^{ 5 } } \nonumber \\
&+\frac { { m }^{ 2 }C }{ 32{ \omega  }_{ k }^{ 5 } } \left( D'''+4D'D+3{ D' }^{ 2 }+6D'D^{ 2 }+D^{ 4 } \right) \nonumber \\
&-\frac { { m }^{ 4 }{ C }^{ 2 } }{ 128{ \omega  }_{ k }^{ 7 } } \left( 28D''D+19{ D' }^{ 2 }+122{ D' }^{ 2 }+47D^{ 4 } \right)  \nonumber \\ 
&+\frac { { 221m }^{ 6 }{ C }^{ 3 } }{ 256{ \omega  }_{ k }^{ 9 } } \left( D'D^{ 2 }+D^{ 4 } \right) 
-\frac { 1105{ m }^{ 8 }{ C }^{ 4 }{ D }^{ 4 } }{ 2048{ \omega  }_{ k }^{ 11 } }+\cdots \label{eq:oekgkf},
\end{align}
where $D=C'/C$. 
Ref.\cite{Parker:1974qw} argued that the mode function with $\alpha_{ k }=1$ and $\beta_{ k }=0$
is a reasonable choice for a sufficiently slow and smooth background,
\begin{align}
{ \delta\chi  }\left( \eta,k  \right)=\frac { 1 }{ \sqrt { 2{ W }_{ k }\left( \eta  \right) C\left( \eta  \right)  }  } 
\cdot e^{-i\int { { W }_{ k }\left( \eta  \right)\, d\eta  }},
\end{align}
which defines so-called adiabatic vacuum $\left| 0 \right>_{\rm A} $ annihilated by all the operators ${ a }_{ k }$,
\begin{align}
{ a }_{ k } \left| 0 \right>_{\rm A}=0.
\end{align}
Note that for a static flat background the mode function ${ \delta\chi  }\left( \eta,k  \right)$ becomes
a positive frequency solution with no adiabatic terms ${ W }_{ k }={ \omega  }_{ k }$
and the adiabatic vacuum reduces to the ordinary Minkowski vacuum $\left| 0 \right>_{\rm M}$. For massless conformal coupled fields 
there is no quantum particle creation 
with respect to the adiabatic vacuum $\left| 0 \right>_{\rm A} $.
However, if we allow the vacuum transition from the Minkowski vacuum $\left| 0 \right>_{\rm M}$ to the 
adiabatic vacuum $\left| 0 \right>_{\rm A} $, 
we can recognize that the background space produces 
massless particles and interpret
the anomaly term of Eq.~(\ref{eq:trace-anomaly}) 
as a contribution of the gravitational particle creations.
Note that the physical interpretation of the quantum conformal anomaly
is non-trivial, but it
affects the dynamics of the spacetime~\cite{Starobinsky:1980te}.

\subsection{Renormalized energy momentum tensor and quantum conformal anomaly}
\label{sec:conformal}
The vacuum expectation values $\left< { T }_{ \mu \nu  } \right>$ of the energy momentum tensor 
for the mode function ${ \delta\chi  }\left(\eta,k \right)$ are given by
\begin{align}
\begin{split}
\left< { T }_{ 00 } \right> &=\frac { 1 }{ 4{ \pi  }^{ 2 }C\left( \eta  \right)  } \int { dk{ k }^{ 2 } } \biggl[ { \left| \delta { \chi ' }_{ k } \right|  }^{ 2 }
+{ \omega  }_{ k }^{ 2 }{ \left| \delta { \chi  }_{ k } \right|  }^{ 2 } \biggr], \\
\left< T_{\phantom{\mu}\mu}^{\mu} \right> &=\frac { 1 }{ 2{ \pi  }^{ 2 }{ C }^2\left( \eta  \right)  } \int { dk{ k }^{ 2 } } 
 \biggl[ C m^{2}{ \left| \delta { \chi  }_{ k } \right|  }^{ 2 }  \biggr],
\end{split}
\end{align}
which satisfy the relation ${ T }_{ ii } = 1/3\left(  { T }_{ 00 }-C\left( \eta  \right)T_{\phantom{\mu}\mu}^{\mu}\right)$.
The energy momentum tensor express the energy density $\rho=\left< { T }_{ 00 } \right>/C\left( \eta  \right)$ and the 
pressure $p=\left< { T }_{ ii } \right>/C\left( \eta  \right)$.
The vacuum expectation values $\left< { T }_{ \mu \nu  } \right>$ of the energy momentum tensor
are written by the adiabatic approximation~\cite{Bunch:1980vc}:
\begin{align}
\begin{split}
\left< { T }_{ 00 } \right> &=\frac { 1 }{ 8{ \pi  }^{ 2 }C\left( \eta  \right)  } \int { dk{ k }^{ 2 } } \biggl[ 
2{ \omega  }_{ k }+\frac{C^{2}m^{4}D^{2}}{16{ \omega  }_{ k }^{ 5}}
-\frac{C^{2}m^{4}}{64{ \omega  }_{ k }^{ 7}}\left(2D''D-D'^{2}+4D'D^{2}+D^{4}\right) \\ &  
+\frac{7C^{3}m^{6}}{64{ \omega  }_{ k }^{ 9}}\left(D'D^{2}+D^{4}\right)-\frac{105C^{4}m^{8}D^{4}}{1024{ \omega  }_{ k }^{11}}  \biggr], \\  
\left< T_{\phantom{\mu}\mu}^{\mu} \right> &=\frac { 1 }{ 4{ \pi  }^{ 2 }C^{2}\left( \eta  \right)  } \int { dk{ k }^{ 2 } } \biggl[ 
\frac{Cm^{2}}{{ \omega  }_{ k }}+\frac{C^{2}m^{4}}{8{ \omega  }_{ k }^{5}}\left(D'+D^{2}\right)-\frac{5C^{3}m^{6}D^{2}}{32{ \omega  }_{ k }^{7}} \\ &
-\frac{C^{2}m^{4}}{32{ \omega  }_{ k }^{7}}\left(D'''+4D''D+3D'^{2}+6D'D^{2}+D^{4}\right)  \\ &
+\frac{C^{3}m^{6}}{128{ \omega  }_{ k }^{9}}\left(28D''D+21D'^{2}+126D'D^{2}+49D^{4}\right)
 -\frac{231C^{4}m^{8}}{256{ \omega  }_{ k }^{11}}\left(D'D^{2}+D^{4}\right)  \\ &
+\frac{1155C^{5}m^{10}D^{4}}{2048{ \omega  }_{ k }^{13}}\biggr].
\end{split}
\end{align}
In order to deal with the renormalization of the quantum energy momentum tensor
$\left< { T }_{ \mu \nu  } \right>$ in curved spacetime,
the adiabatic expansion must be performed up to the fourth order~\cite{Bunch:1980vc}.
However, the high-order adiabatic terms are finite and 
the divergences of the energy momentum tensor originate from the lowest-order adiabatic term
\begin{align}
\left< { T }_{ 00 } \right>_{\rm diverge}=\frac { 1 }{ 4{ \pi  }^{ 2 }C\left( \eta  \right)  } \int { dk{ k }^{ 2 } }{ \omega  }_{ k },
\end{align}
Adopting the dimensional regularization
\footnote{
The divergent momentum integrals are simplified as
\begin{align*}
I\left( 0,n \right) =\int { \frac { { d }^{ 3 }k }{ { \left( 2\pi  \right)  }^{ 3 } } \frac { 1 }{ { \omega  }_{ k }^{ n } } 
=\int { \frac { { d }^{ 3 }k }{ { \left( 2\pi  \right)  }^{ 3 } } \frac { 1 }{ \left( { k }^{ 2 }+{ a }^{ 2 }{ m }^{ 2 } \right) ^{ n/2 } } }  }.
\end{align*}
We regulate these integrals of the spatial dimensions $ 3-2\epsilon $ as 
\begin{align*}
I\left( \epsilon,n \right) =\int { \frac { { d }^{ 3-2\epsilon }k }{ { \left( 2\pi  \right)  }^{3-2\epsilon } } }
\frac { \left(a\mu\right)^{2\epsilon} }{ { \omega  }_{ k }^{ n } }=\frac { { \left( am \right)  }^{ 3-n }}
{ 8{ \pi  }^{ 3/2 } }
\frac { \Gamma \left( \epsilon -\frac { 3-n }{ 2 }  \right)  }{ \Gamma \left( \frac { n }{ 2 }  \right)  }
{ \left( \frac { 4\pi { \mu  }^{ 2 } }{ { m }^{ 2 } }  \right)  }^{ \epsilon  }=
\frac { { \left( am \right)  }^{ 3-n }}
{ 8{ \pi  }^{ 3/2 } }
\frac { \Gamma \left( \epsilon -\frac { 3-n }{ 2 }  \right)  }{ \Gamma \left( \frac { n }{ 2 }  \right)  }
{ \left( 1+\epsilon \cdot \ln {\frac { 4\pi { \mu  }^{ 2 } }{ { m }^{ 2 } } }  \right)  }.
\end{align*}
}
the lowest-order adiabatic term of the energy momentum tensor
can be regularized as follows:
\begin{align}
\begin{split}
\left< { T }_{ 00 } \right>_{\rm diverge} =-\frac { { m }^{ 4 }C}{ 64{ \pi  }^{ 2 } } \left[ \frac { 1 }{ \epsilon  } 
+\frac { 3 }{ 2 } -\gamma +\ln { 4\pi  } +\ln { \frac { { \mu  }^{ 2 } }{ { m }^{ 2 } }  }  \right].
\end{split}
\end{align}
The regularized vacuum expectation values of $\left< { T }_{ 00 } \right>$ the energy momentum tensor in curved spacetime
are written as 
\begin{align}
\begin{split}
\left< { T }_{ 00 } \right>_{\rm regularized } &=-\frac { { m }^{ 4 }C}{ 64{ \pi  }^{ 2 } } \left[ \frac { 1 }{ \epsilon  } 
+\frac { 3 }{ 2 } -\gamma +\ln { 4\pi  } +\ln { \frac { { \mu  }^{ 2 } }{ { m }^{ 2 } }  }  \right] \\
&+\frac { { m }^{ 2 }{ D }^{ 2 } }{ 384{ \pi  }^{ 2 } }-\frac { 1 }{ 2880{ \pi  }^{ 2 }{ C }} 
\left( \frac { 3 }{ 2 } D''D-\frac { 3 }{ 4 } { D' }^{ 2 }-\frac { 3 }{ 8 } { D }^{ 4 } \right),
\end{split}
\end{align}
where the high-order adiabatic terms are finite and correspond to the
gravitational particle creation effects in curved spacetime.
The lowest-order adiabatic term has the same UV divergent structure as the expressions of
the Minkowski vacuum $\left| 0 \right>_{\rm M}$ and we can remove these contributions 
renormalizing the gravitational coupling constants in the Einstein equation.
Thus, the adiabatic (WKB) approximation is the powerful method to obtain the renormalized 
energy momentum tensor.

The general Einstein's equations with the vacuum expectation values 
 of the energy momentum tensor are given by
\begin{align}\label{eq:Einstein's equations}
\frac { 1 }{ 8\pi { G }_{ N } } { G }_{ 00  }+{ \rho  }_{ \Lambda  } 
 { g }_{ 00 }+{ a }_{ 1 }{ H }_{ 00  }^{ \left( 1 \right)  }
+{ a }_{ 2 }{ H }_{ 00  }^{ \left( 2 \right)  }+{ a }_{ 3 }{ H }_{ 00  }^{ \left( 3 \right)  }
=\left< { T }_{ 00 } \right>,
\end{align}where:
\begin{align*}
\begin{split}
H^{(1)}_{\mu\nu} &\equiv 2\nabla_\nu \nabla_\mu R -2g_{\mu\nu}\Box R
 - {1\over 2}g_{\mu\nu} R^2 +2R R_{\mu\nu},   \\
H^{(2)}_{\mu\nu} &\equiv
2\nabla_\alpha \nabla_\nu R_\mu^\alpha - \Box R_{\mu\nu} -{1\over 2}g_{\mu\nu}\Box R
  -{1\over 2}g_{\mu\nu} R_{\alpha\beta}R^{\alpha\beta} 
   +2R_\mu^\rho R_{\rho\nu},     \\
H^{(3)}_{\mu\nu} &\equiv -  H^{(1)}_{\mu\nu} + 4H^{(2)}_{\mu\nu},
\end{split}
\end{align*}
where ${ G }_{ \mu \nu  }={ R }_{ \mu \nu  }-\frac { 1 }{ 2 }R{ g }_{ \mu \nu  }$ is the Einstein tensor and 
${ \rho  }_{ \Lambda  } \equiv \Lambda /{8\pi { G }_{ N }}$.
The unphysical divergences of the energy momentum tensor 
are absorbed by the counter terms $\delta{ G }_{ N }$, $\delta{ \rho  }_{ \Lambda  }$, 
$\delta{ a }_{ 1,2,3}$ of the gravitational couplings as follows~\cite{Kohri:2016lsj}:
\begin{align}
\begin{split}
 \left< { T }_{ 00  } \right>_{\rm diverge}&=
\frac { 1 }{ 8\pi \delta{ G }_{ N } } { G }_{ 00  }+\delta{ \rho  }_{ \Lambda  }{ g }_{ 00  }+\delta{ a }_{ 1 }{ H }_{ 00  }^{ \left( 1 \right)  }
+\delta{ a }_{ 2 }{ H }_{ 00  }^{ \left( 2 \right)  } +\delta{ a }_{ 3 }{ H }_{ 00  }^{ \left( 3 \right)  } \\ &
=\frac { 1 }{ 8\pi \delta{ G }_{ N } }\left( \frac { 3 }{ 4 }{ D }^{ 2 }   \right) + \delta{ \rho  }_{ \Lambda  } \left( -C \right)
+ \delta{ a }_{ 1 } \left( \frac{72D''D-36D'^{2}-27D^{4}}{8C} \right)+\cdots.
\end{split}
\end{align}
Thus, the renormalized energy momentum tensor $\left< { T }_{ 00 } \right>_{\rm ren}$ can be written as
\begin{align}\label{eq:tugdfsdg}
\left< { T }_{ 00 } \right>_{\rm ren}
= \frac { { m }^{ 4 }C}{ 64{ \pi  }^{ 2 } }
\left(\ln { \frac { { m  }^{ 2 } }{ { \mu }^{ 2 } }  }-\frac{3}{2}\right) 
+\frac { { m }^{ 2 }{ D }^{ 2 } }{ 384{ \pi  }^{ 2 }}-\frac { 1 }{ 2880{ \pi  }^{ 2 }{ C }} 
\left( \frac { 3 }{ 2 } D''D-\frac { 3 }{ 4 } { D' }^{ 2 }-\frac { 3 }{ 8 } { D }^{ 4 } \right).
\end{align}
The first part originates from the lowest-order adiabatic term of $\left< { T }_{ 00 } \right>$ which is 
renormalized by the cosmological constant term ${ \rho  }_{ \Lambda  }$.
On the other hand, the second parts originating from the high-order adiabatic terms
express the quantum gravitational contributions on curved spacetime.
Although there are no ambiguity about existence of the renormalized energy momentum tensor 
$\left< { T }_{ \mu\nu } \right>_{\rm ren}$, 
there is some ambiguity about the renormalized expressions 
$\left< { T }_{ \mu\nu } \right>_{\rm ren}$.
The ambiguity in various methods is closely related
with what to take the vacuum and 
how to determine gravitational particle productions
rather than the regularization of the UV divergences~\cite{Birrell:1982ix}.
Thus, taking the adiabatic approximation for the mode function and 
the adiabatic vacuum $\left| 0 \right>_{\rm A} $ 
gives a definite expression for the renormalized energy momentum tensor 
$\left< { T }_{ \mu\nu } \right>_{\rm ren}$.

Let us consider the trace of the vacuum expectation values 
$\left< { T }_{ \mu \nu  } \right>$ of the energy momentum tensor,
\begin{align}
\begin{split}
\left< T_{\phantom{\mu}\mu}^{\mu}\right> &=\frac { m^{2} }{ 4{ \pi  }^{ 2 }C\left( \eta  \right)  } \int { dk{ k }^{ 2 } } \biggl[ 
\frac{1}{{ \omega  }_{ k }}+\frac{Cm^{2}}{8{ \omega  }_{ k }^{5}}\left(D'+D^{2}\right)-\frac{5C^{2}m^{4}D^{2}}{32{ \omega  }_{ k }^{7}} \\ &
-\frac{C m^{2}}{32{ \omega  }_{ k }^{7}}\left(D'''+4D''D+3D'^{2}+6D'D^{2}+D^{4}\right)  \\ &
+\frac{C^{2}m^{4}}{128{ \omega  }_{ k }^{9}}\left(28D''D+21D'^{2}+126D'D^{2}+49D^{4}\right)
 -\frac{231C^{3}m^{6}}{256{ \omega  }_{ k }^{11}}\left(D'D^{2}+D^{4}\right)  \\ &
+\frac{1155C^{4}m^{8}D^{4}}{2048{ \omega  }_{ k }^{13}}\biggr] \\
&=-\frac { { m }^{ 4 }}{ 32{ \pi  }^{ 2 }C } \left[ \frac { 1 }{ \epsilon  } 
+1 -\gamma +\ln { 4\pi  } +\ln { \frac { { \mu  }^{ 2 } }{ { m }^{ 2 } }  }  \right] 
+\frac { { m }^{ 2 }{ D }^{ 2 } }{ 192{ \pi  }^{ 2 }C }\left(2D'+D^2\right) -\frac { 1 }{ 960{ \pi  }^{ 2 }{ C }^{ 2 } } \left( D'''-D'{ D }^{ 2 } \right) .
\end{split}
\end{align}
The renormalized trace $\left< T_{\phantom{\mu}\mu}^{\mu}\right>_{\rm ren}$ 
of the quantum energy momentum tensor is given by
\begin{align}\label{eq:trace-anomaly}
\begin{split}
\left< T_{\phantom{\mu}\mu}^{\mu}\right>_{\rm ren} 
=\frac { { m }^{ 4 }}{ 32{ \pi  }^{ 2 }C } \left( \ln { \frac { { m  }^{ 2 } }{ { \mu }^{ 2 } }  } -1 \right)
+\frac { { m }^{ 2 }{ D }^{ 2 } }{ 192{ \pi  }^{ 2 }C }\left(2D'+D^2\right) 
-\frac { 1 }{ 960{ \pi  }^{ 2 }{ C }^{ 2 } } \left( D'''-D'{ D }^{ 2 } \right) .
\end{split}
\end{align}
The anomaly term of Eq.~(\ref{eq:trace-anomaly}) is consistent with 
using dimensional regularization~\cite{Deser:1976yx,Duff:1977ay} and is equal to $a_2(x)/16 \pi^2$~\cite{Buchbinder:1992rb}
where $a_2(x)$ is a coefficient of the DeWitt-Schwinger formalism. 
The conformal anomaly is given by the massless limit of Eq.~(\ref{eq:trace-anomaly}) 
\begin{align}\label{eq:conformal-anomaly}
\begin{split}
\left< T_{\phantom{\mu}\mu}^{\mu}\right>_{\rm anomaly} =\lim _{ m\rightarrow 0 }{ \left< T_{\phantom{\mu}\mu}^{\mu}\right>_{\rm ren}  } &
=-\frac { 1 }{ 960{ \pi  }^{ 2 }{ C }^{ 2 } } \left( D'''-D'{ D }^{ 2 } \right) \\ &
=-\frac { 1 }{ 2880{ \pi  }^{ 2 } } \left[ \left( { R }_{ \mu \nu  }{ R }^{ \mu \nu  }-\frac { 1 }{ 3 } { R }^{ 2 } \right) +\Box R \right] \\ &
=\frac{1}{360(4\pi)^{2}}\left(E-\frac{2}{3}\Box R\right) + \frac{-1}{270(4\pi)^{2}}\Box R \\ &
=\frac{1}{360(4\pi)^{2}}E - \frac{1}{180(4\pi)^{2}}\Box R  ,
\end{split}
\end{align}
where we used the relations for the flat FLRW spacetime 
\begin{align}
C_{\mu\nu\kappa\lambda} C^{\mu\nu\kappa\lambda} =
R_{\mu\nu\kappa\lambda} R^{\mu\nu\kappa\lambda}- 2 R_{\mu\nu}
R^{\mu\nu} + \frac{1}{3} R^{2}=0.
\end{align}
Generally, the renormalized expressions for conformal anomaly
includes regularization-scheme dependent parts.
However, the adiabatic approximation systematically removes UV divergences and 
gives a definite expressions.
In this sense, the conformal anomaly of Eq.~(\ref{eq:conformal-anomaly})
has no ambiguity if the adiabatic vacuum $\left| 0 \right>_{\rm A} $ is appropriate
for cosmological situation and we can interpret the conformal anomaly 
as the quantum gravitational effects. 
Note that Eq.~(\ref{eq:conformal-anomaly}) is consistent with the results 
from the effective action $\Gamma_{\rm eff} \left[g_{\mu\nu}\right]$ of gravity~\cite{Buchbinder:1992rb}
in the dimensional regularization.
On the other hand the adiabatic regularization leads to the wrong sign of the 
conformal anomaly~\cite{Bunch:1980vc}.
That originates from the subtraction of the adiabatic mode function~\cite{Bunch:1980vc},
\begin{align}\label{eq:ren-anomaly}
\begin{split}
\left< T_{\phantom{\mu}\mu}^{\mu}\right>_{\rm anomaly} &=
\left< T_{\phantom{\mu}\mu}^{\mu}\right>-\lim _{ m\rightarrow 0 }{ \left< T_{\phantom{\mu}\mu}^{\mu}\right>} \\ &
=\frac { 1 }{ 960{ \pi  }^{ 2 }{ C }^{ 2 } } \left( D'''-D'{ D }^{ 2 } \right) \\ &
=-\frac{1}{360(4\pi)^{2}}E + \frac{1}{180(4\pi)^{2}}\Box R.
\end{split}
\end{align}
The adiabatic regularization~\cite{Bunch:1980vc,Parker:1974qw,Fulling:1974pu,Fulling:1974zr,Anderson:1987yt,Kohri:2016lsj}
is the powerful method to obtain the renormalized 
energy momentum tensor, but it is not appropriate for deriving a concrete expression of the conformal anomaly
since the mode functions of the conformal fields already satisfy the adiabatic condition~\cite{Birrell:1982ix}.
By using the adiabatic (WKB) approximation for massless fermions~\cite{Landete:2013lpa,delRio:2014cha}
we obtain the following expression for the conformal anomaly, 
\begin{align}\label{eq:conformal-anomaly-f}
\begin{split}
\left< T_{\phantom{\mu}\mu}^{\mu}\right>_{\rm anomaly}^{\rm fermion}
&=-\frac { 1 }{ 2880{ \pi  }^{ 2 } } \left[ 11\left( { R }_{ \mu \nu  }{ R }^{ \mu \nu  }-\frac { 1 }{ 3 } { R }^{ 2 } \right) + 6\, \Box R \right] \\ &
=\frac{11}{360(4\pi)^{2}}E - \frac{6}{180(4\pi)^{2}}\Box R.
\end{split}
\end{align}
The conformal anomaly for the gauge field in adiabatic expansion
is given by~\cite{Chu:2016kwv}:
\begin{align}\label{eq:conformal-anomaly-g}
\begin{split}
\left< T_{\phantom{\mu}\mu}^{\mu}\right>_{\rm anomaly}^{\rm gauge\, boson}
&=-\frac { 1 }{ 2880{ \pi  }^{ 2 } } \left[ 62\left( { R }_{ \mu \nu  }{ R }^{ \mu \nu  }-\frac { 1 }{ 3 } { R }^{ 2 } \right) 
- \left(18+15 \log \xi \right) \Box R \right] \\ &
=\frac{62}{360(4\pi)^{2}}E + \frac{\left(18+15 \log \xi \right)}{180(4\pi)^{2}}\Box R.  
\end{split}
\end{align}
where $\xi$ is a gauge fixing parameter defined by the covariant gauge fixing term~\cite{Chu:2016kwv}:
\begin{align}
\mathcal{ L  }_{ \rm gf }=-\frac { \sqrt { -g }  }{ 2\xi  } { \left( { \nabla  }^{ \mu  }{ A }_{ \mu  } \right)  }^{ 2 }.
\end{align}
The gauge dependence of Eq.~(\ref{eq:conformal-anomaly-g}) exists in the DeWitt-Schwinger expansion
formalism~\cite{Endo:1984sz,Toms:2014tia,Vieira:2015oka}. The adiabatic expansion or regularization reproduce the 
gauge dependence of the $\Box R$ term which has also the regularization-scheme dependence.
However, the gauge fixing parameter can be removed by the the gravitational coupling constants
in the Einstein equation, and therefore, from here we drop the gauge fixing parameter $\xi$.
Now, we point out that the adiabatic expressions for the conformal anomaly precisely matches the expression derived by
the effective action $\Gamma_{\rm eff} \left[g_{\mu\nu}\right]$ with the dimensional regularization~\cite{Buchbinder:1992rb}.

\section{De Sitter spacetime instability from conformal anomaly}
\label{sec:instability}
In this section we discuss the de Sitter instability from quantum backreaction 
involving the conformal anomaly~\cite{Capper:1974ic,Deser:1976yx}. 
The nontrivial effect of the conformal anomaly on the spacetime 
has been widely discussed in literature.
For instance, the FLRW spacetime with the conformal anomaly can reduce a big bang singularity~\cite{Nojiri:2004ip,Awad:2015syb}.
The genuine non-locality of the conformal anomaly homogeneously influences the dynamics of the universe.
Thus, it has been argued that the conformal anomaly could have an effect on the dark energy or 
the cosmological constant problem~\cite{Tomboulis:1988gw,Antoniadis:1991fa,Antoniadis:1998fi,
Schutzhold:2002pr,Antoniadis:2006wq,Bilic:2007gr,Koksma:2008jn}.
And also, the conformal anomaly can provide the quantum source of the inflation. 
Originally, Starobinsky~\cite{Starobinsky:1980te} introduced that the conformal anomaly 
lead to the de Sitter solution of the spacetime.
 Subsequently, this theory has been improved as the anomaly induced inflation~\cite{Hawking:2000bb,Shapiro:2001rh,
Pelinson:2002ef,Netto:2015cba}.
Here, we focus on the stability of de Sitter spacetime involving the conformal anomaly 
and clearly show that higher-derivative terms from the conformal anomaly
destabilize the background spacetime.
Our results reproduces some results of the literature~\cite{Starobinsky:1980te,Hawking:2000bb,Shapiro:2001rh,
Pelinson:2002ef,Netto:2015cba}, but we clearly show that
the cosmological constant does not determine the ultimate dynamics
of the spacetime except for the initial behavior.
The results suggest the de Sitter spacetime instability from the semiclassical gravity 
and the cosmic inflation should not last long for the specific conditions.
The cosmological application of the de Sitter instability will be discussed 
in the later Section~\ref{sec:cosmic_inflation}.

Our analysis is based on the semiclassical approach of the gravity.
As previous discussed in Section~\ref{sec:backreaction}, this formalism 
includes quantum backreaction on the spacetime
and can be regarded as the low-energy effective theory of QG.
The semiclassical gravity is described by the effective Einstein's equations of Eq.~(\ref{eq:Einstein's equations}) with 
the renormalized vacuum expectation values $\left< { T }_{ \mu \nu  } \right>_{\rm ren}$ of 
the energy momentum tensor,
\begin{align}\label{Einstein's equations}
\frac { 1 }{ 8\pi { G }_{ N } }\left(
{ R }_{ \mu \nu  }-\frac { 1 }{ 2 }R{ g }_{ \mu \nu  }+{ \Lambda  }
 { g }_{ \mu \nu  }\right)+{ a }_{ 1 }{ H }_{ \mu \nu  }^{ \left( 1 \right)  }
{ +a }_{ 2 }{ H }_{ \mu \nu  }^{ \left( 2 \right)  }{ +a }_{ 3 }{ H }_{ \mu \nu  }^{ \left( 3 \right)  }=\left< { T }_{ \mu\nu } \right>_{\rm ren}.
\end{align}
The trace of the renormalized energy momentum tensor $\left< { T }_{ \mu \nu  } \right>_{\rm ren}$
lead to the conformal anomaly. Here we assume that there exits 
a number of free conformally invariant fields in the universe. 
The conformal anomaly contributions of Eq.~(\ref{eq:conformal-anomaly})
is one-loop quantum correction and therefore the description based on Eq.~(\ref{eq:conformal-anomaly})
is physically exact as long as $R_{\mu\nu\kappa\lambda} R^{\mu\nu\kappa\lambda} \ll M_{\rm P}^4 $.
If this is not satisfied the higher-loop corrections 
of the matter fields or the graviton cannot be neglected. 
However, the one-loop expression is sufficient to estimate the behavior of the system approximately.
Furthermore, we can neglect the higher-loop corrections 
in a specific theory like $\mathcal{N}=4$ Super Yang Mills (SYM) theory  
where the conformal anomaly is one-loop exact~\cite{Osborn:1993cr,Anselmi:1996mq,Anselmi:1997am,Anselmi:1996dd}.
For this reason, our purpose could be sufficiently achieved even in the one-loop expression.
The general form of the conformal anomaly for four dimensions is given by 
\begin{align}\label{TraceAnomaly1}
\left\langle
T_{\phantom{\mu}\mu}^{\mu}\right\rangle_{\rm anomaly} = b F + b' E + c\, \Box R \,.
\end{align}
The dimensionless parameters $b$, $b'$ and $c$ of
the adiabatic (WKB) approximation are given by:
\begin{subequations}
\label{TraceAnomaly3}
\begin{align}
b &= -\frac{1}{120(4\pi)^{2}}\left( N_{S}+6N_{F}+ 12 N_{G}\right)
 \label{TraceAnomaly3a}, \\
b' &= \frac{1}{360(4\pi)^{2}}\left( N_{S}+\frac{11}{2}N_{F}+ 62 N_{G}\right), \\
c &= -\frac{1}{180(4\pi)^{2}}\left( N_{S}+6 N_{F}-18 N_{G}\right),
\label{TraceAnomaly3b}
\end{align}
\end{subequations}
where we consider $N_{S}$ scalars (spin-0), $N_{F}$ Dirac fermions (spin-1/2) 
and $N_{G}$ abelian gauge (spin-1) fields. 
These parameters $b$, $b'$ and $c$ can be rewritten in concrete models.
Now we disregard all spin-2 and spin-3/2 degrees of freedom
since gravity is treated as purely classical background.
For the Minimal Supersymmetric Standard Model (MSSM), 
we can take the following values: $N_{S}=104$, $N_{F}=32$ and $N_{G}=12$.
For the Standard Model context, we take the following values: $N_{S}=4$,
$N_{F}=24$ and $N_{G}=12$ where we consider the existence of the right-handed neutrinos. 
Finally, the massless particle in current universe 
is only photon, thus, $N_{V}=1$, $N_{S}=0$ and $N_{G}=0$.

The effective Einstein's equations of Eq.~(\ref{Einstein's equations}) 
describes the dynamics of the background spacetime or the Universe. 
The differential equation derived from the Einstein field equations 
with $\left\langle
T_{\phantom{\mu}\mu}^{\mu}\right\rangle_{\rm anomaly}$ can be given as follows~\cite{Starobinsky:1980te,Hawking:2000bb,Shapiro:2001rh,
Pelinson:2002ef,Netto:2015cba}:
\begin{align}
\label{eq:T00}
\frac{2\, {\stackrel{.}{a}}\, {\stackrel{...}{a}} }{a^2}
-\frac{{\stackrel{..}{a}}^2}{a^2}+
\frac{2\, {\stackrel{..}{a}} \, {\stackrel{.}{a}}^2}{a^3}
-\Big(3+\frac{2 b'}{c}\Big)\frac{{\stackrel{.}{a}}^4}{a^4}-\frac{1}{8\pi c \,  G }\frac{\Lambda}{3}
+\frac{1}{8\pi c \,  G} \, \frac{{\stackrel{.}{a}}^2}{a^2}
\,=\, 0\, ,
\end{align}
where we drop the gravitational derivative couplings ${ a }_{ 1,2,3}$ for simplicity.
Note that a system are generally destabilized by
such higher derivative terms and the initial conditions must be highly fine-tuned. 
Formally, this instability is known as the Ostrogradsky theorem~\cite{Woodard:2006nt}.
Now let us rewrite the above differential equation with respect to the Hubble parameter,
\begin{align}\label{eq:Hubble-equation}
6{ H }^{ 2 }\dot { H } +2H\ddot { H } -{ \dot { H }  }^{ 2 }-\frac{2b'}{c}{ H }^{ 4 }-\frac{1}{8\pi c \,  G }\frac{\Lambda}{3}
+\frac{1}{8\pi c \,  G} \, { H }^{ 2 }=0.
\end{align}
To capture the asymptotic behavior for the
Hubble parameter in the effective Einstein's equations
we ignore the time derivative term of Eq.~(\ref{eq:Hubble-equation}) as follows:
\begin{align}\label{eq:dHubble-equation}
{ H }^{ 4 }-\frac{1}{16\pi b'\,  G} \, { H }^{ 2 }
+\frac{1}{16\pi b' \,  G }\frac{\Lambda}{3}=0.
\end{align}
Solving Eq.~(\ref{eq:dHubble-equation})
we obtain the asymptotic expressions for the
Hubble parameter as follows:
\begin{align}\label{eq:Asymptote}
H^2=\left(\frac{1}{32\pi b' G } \right)\pm \frac{1}{{4b'}}\sqrt{\left(\frac{1}{8\pi G } \right)^2 -
\left(\frac{1}{8\pi G } \right)\frac{8b'\Lambda}{3}} \, ,
\end{align}
which are completely independent of the coefficient $c$.
Solving Eq.~(\ref{eq:Asymptote}) for the relatively small cosmological constant
$8 b' \Lambda/3 \ll M_{\rm P}$
we get two asymptotic/stable inflationary attractors,
\begin{align}
H_{\mathrm{C}} \simeq \sqrt{\frac{\Lambda}{3}},\quad H_{\mathrm{Q}} \simeq\sqrt{\frac{1}{16\pi b' G }},
\end{align}
where $H_{\mathrm{C}}$ turns out to be the classical de Sitter attractor 
and $H_{\mathrm{Q}}$ is a quantum anomaly driven attractor. 
In the case of zero cosmological constant, the conformal anomaly can
provide us with anomaly-induced inflation scenario
discussed by~\cite{Hawking:2000bb,Shapiro:2001rh,Pelinson:2002ef,Netto:2015cba}.

\begin{figure*}[t]
        \begin{tabular}{cc}
	\begin{minipage}{0.5\hsize}
		\centering
		\subfigure[$b'=8/45\pi^{2}$ and $c=-1/36\pi^{2}$]{
		\includegraphics[width=75mm]{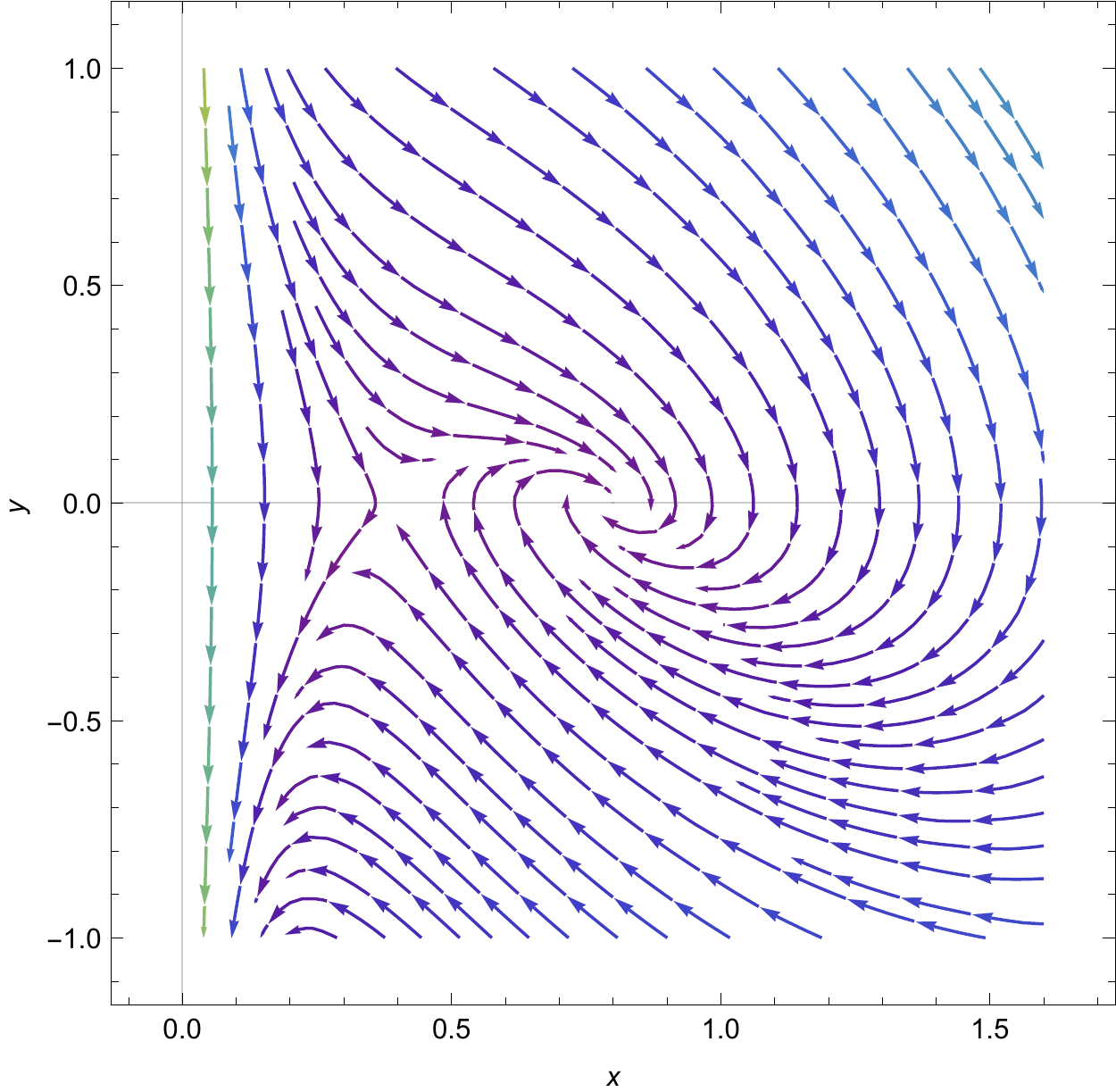}
		\label{fig:MSSMs}}\end{minipage}
\begin{minipage}{0.5\hsize}
		\centering
		\subfigure[$b'=11/72\pi^{2}$ and $c=17/720\pi^{2}$]{
		\includegraphics[width=75mm]{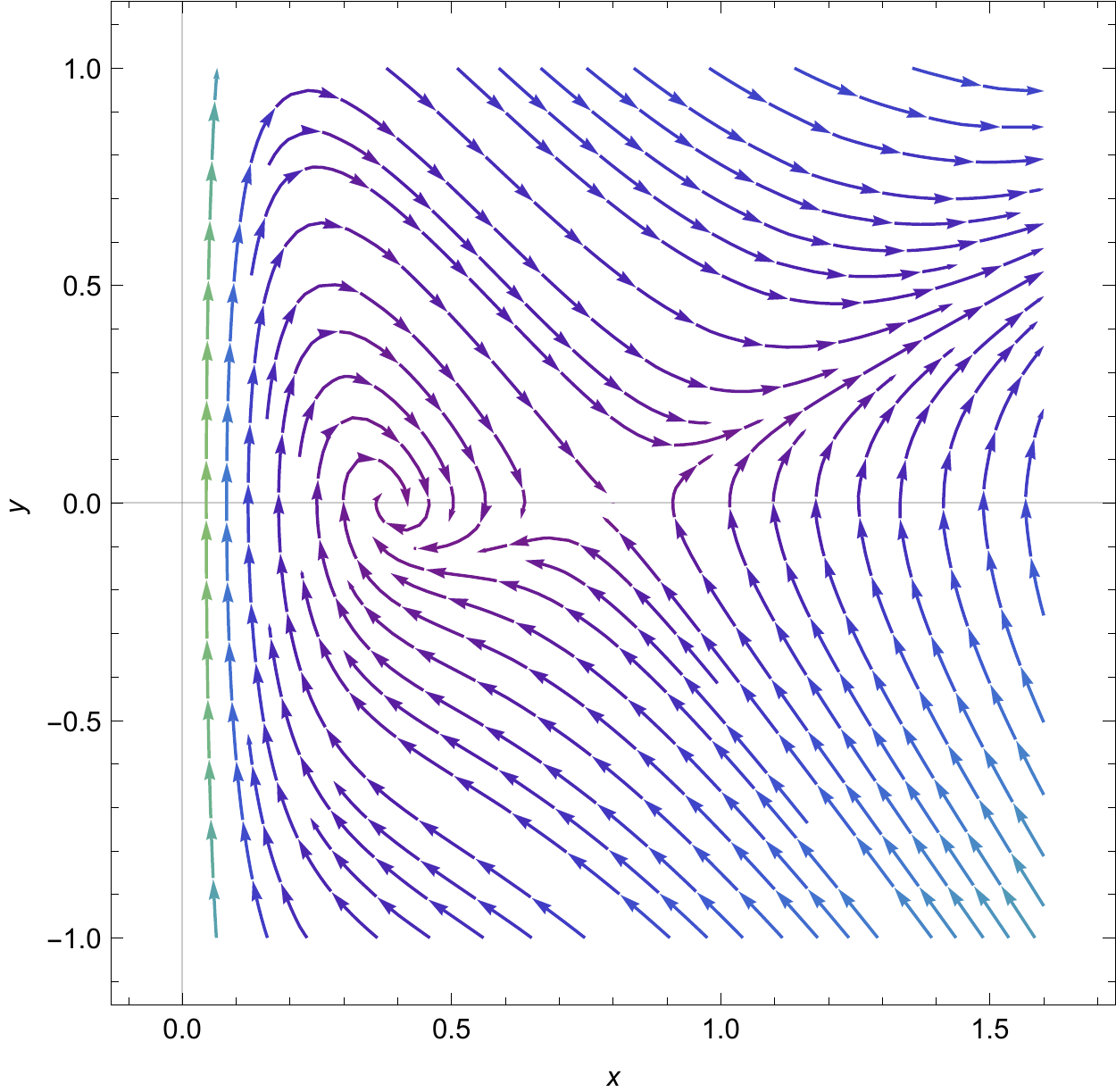}
		\label{fig:SMs}}
\end{minipage}
\end{tabular}
\caption{Phase diagram on Eq.~(\ref{eq:phased_cos}) as a function of $x$ and $y$.
  Left figure shows the Hubble phase diagram with the cosmological constant
  $2 b'\Lambda/3M_{\mathrm{P}}^2 =10^{-0.7}$
  and assume $b'=8/45\pi^{2}$, $c=-1/36\pi^{2}$. 
  On the other hand, 
  right figure shows the Hubble phase diagram with the cosmological constant
  $2 b'\Lambda/3M_{\mathrm{P}}^2 =10^{-0.7}$
  and set $b'=11/72\pi^{2}$, $c=17/720\pi^{2}$.
  The above two critical points correspond to 
  the classical and quantum de Sitter solution with $H_{\mathrm{C}} \simeq \sqrt{ {\Lambda}/{3}}$ and
  $H_{\mathrm{Q}}\simeq  M_{\mathrm{P}}/\sqrt{2 b'}$.
  }\label{Fig:phase diagram_cos11}
\end{figure*}
\begin{figure*}[h]
        \begin{tabular}{cc}
	\begin{minipage}{0.5\hsize}
		\centering
		\subfigure[$b'=8/45\pi^{2}$ and $c=-1/36\pi^{2}$]{
		\includegraphics[width=75mm]{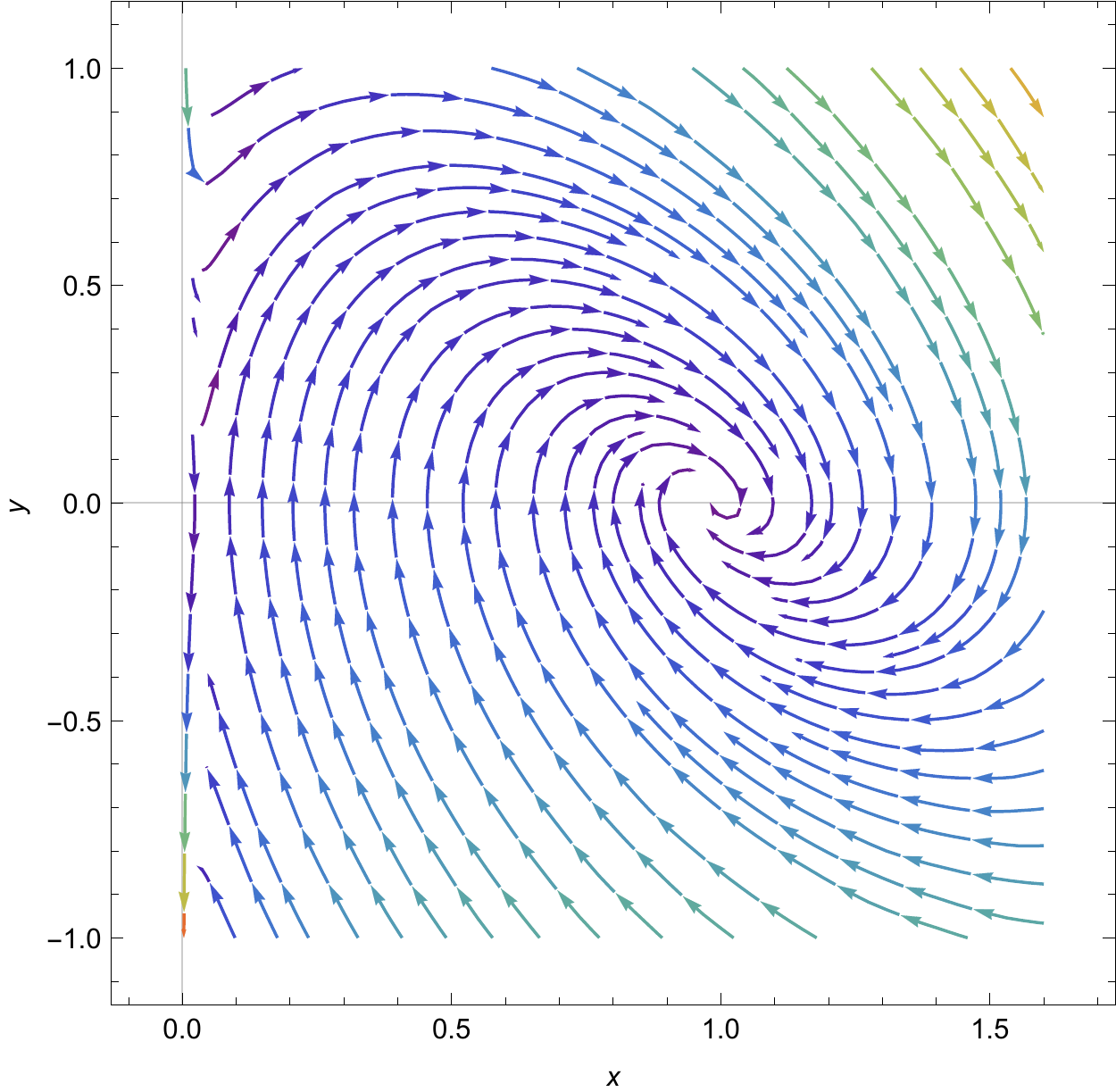}
		\label{fig:MSSMs}}\end{minipage}
\begin{minipage}{0.5\hsize}
		\centering
		\subfigure[$b'=11/72\pi^{2}$ and $c=17/720\pi^{2}$]{
		\includegraphics[width=75mm]{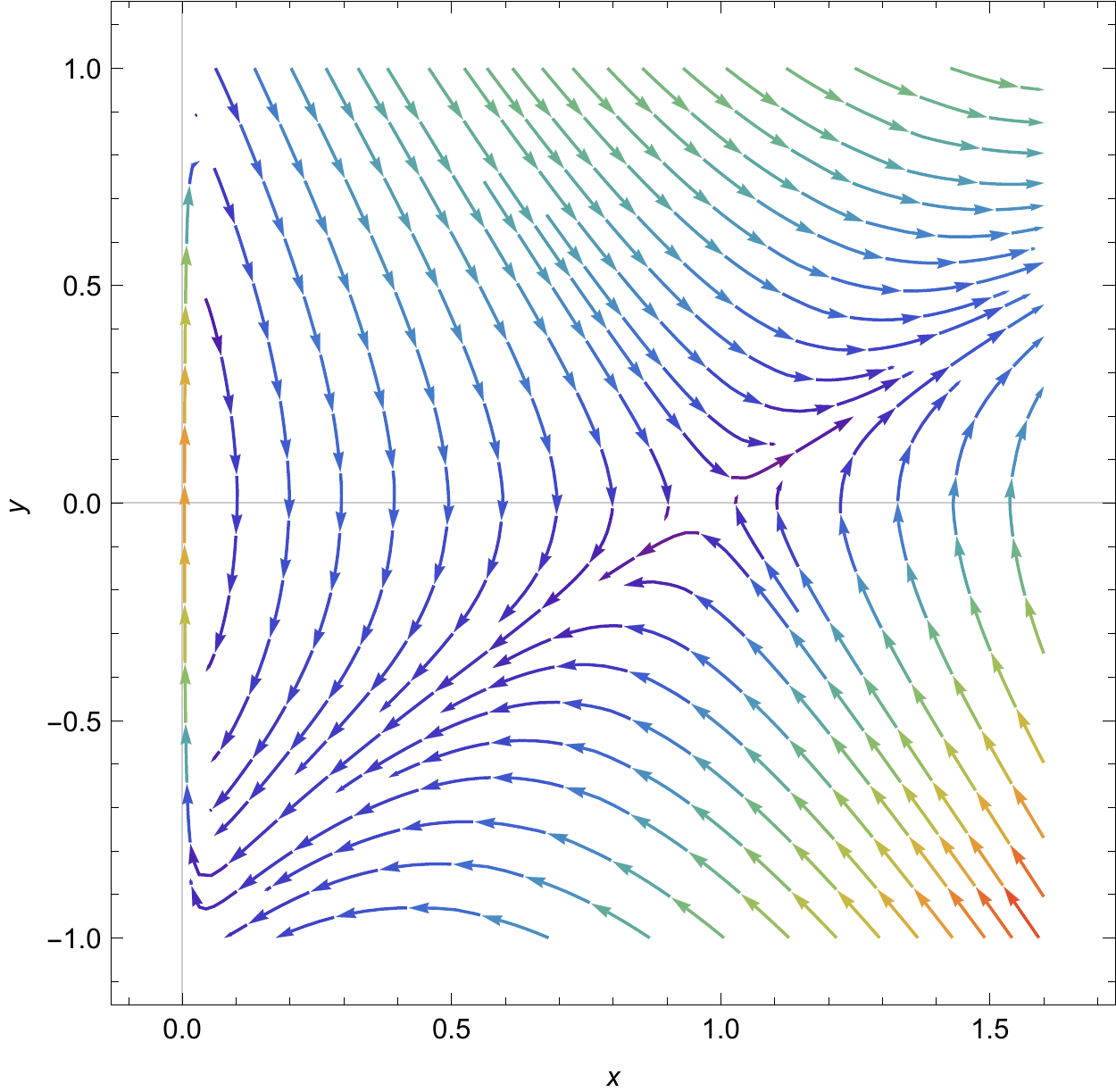}
		\label{fig:SMs}}
\end{minipage}
\end{tabular}
\caption{Phase diagram on Eq.~(\ref{eq:phased_cos}) as a function of $x$ and $y$.  
  Left figure shows the Hubble phase diagram with the cosmological constant
  $2 b'\Lambda/3M_{\mathrm{P}}^2 =10^{-2.0}$
  and assume $b'=8/45\pi^{2}$, $c=-1/36\pi^{2}$. 
  On the other hand, 
  right figure shows the Hubble phase diagram with the cosmological constant
  $2 b'\Lambda/3M_{\mathrm{P}}^2 =10^{-2.0}$
  and set $b'=11/72\pi^{2}$, $c=17/720\pi^{2}$.
  The above two critical points correspond to 
  classical/quantum de Sitter attractor with $H_{\mathrm{C}} \simeq \sqrt{ {\Lambda}/{3}}$ and
  $H_{\mathrm{Q}}\simeq  M_{\mathrm{P}}/\sqrt{2 b'}$.
}\label{Fig:phase diagram_cos12}
\end{figure*}

Next, let us numerically solve Eq.~(\ref{eq:T00}) and investigate the stability of 
de Sitter spacetime involving the conformal anomaly.
The Eq.~(\ref{eq:T00}) and Eq.~(\ref{eq:Hubble-equation}) can be reduced to 
the first-order differential equation~\cite{Starobinsky:1980te, Netto:2015cba}:
\begin{align}
\label{eq:phase}
\frac{dy}{dx}=
\frac{b'(x-x^{-1/3}+\frac{2b'\Lambda}{3M^2_{\rm P}}x^{-5/3})}{6cy} - 1,
\end{align}
by adopting the following variables:
\begin{align}
\nonumber
x= \Big(\frac{H}{H_{\mathrm{Q}}}\Big)^{3/2},\quad
y = \frac{\dot{H}}{2H_{\mathrm{Q}}^{3/2}} \, H^{-1/2},\quad
dt = \frac{dx}{3 H_{\mathrm{Q}}x^{2/3}y},\end{align}
where we introduce $H_{\mathrm{Q}}= M_{\mathrm{P}}/\sqrt{2 b'}$.
Then, the slow-roll parameter can be rewritten with respect to these variables
\begin{align}\label{eq:slow}
\epsilon \equiv -\frac{\dot{H}}{H^{2}} = \frac{-2y}{x}\ll 1.
\end{align}
Solving Eq.~(\ref{eq:phase}) we consider the following two differential equations,
\begin{align}
\label{eq:phased_cos}
\frac{dx}{d \tau}=3 x^{2/3}y, \quad
\frac{dy}{d \tau}=
\frac{b'\, (x^{5/3}-x^{1/3} +\frac{2b'\Lambda}{3M^2_{\rm P}}x^{-1})}{2c} - 3\, x^{2/3}y,
\end{align}
where we set $\tau=t\cdot H_{\mathrm{Q}}$.
For zero cosmological constant case $\Lambda =0$, 
the conformal anomaly provides stable or unstable 
de Sitter attractors~\cite{Starobinsky:1980te,Netto:2015cba}.
For the MSSM particle content where $b'=8/45\pi^{2}$ and $c=-1/36\pi^{2}$, the phase
diagram shows a nonsingular attractor corresponding to the inflationary solution.
This solution reduces the initial or big bang singularity of the spacetime~\cite{Starobinsky:1980te}.
For the SM contents where $b'=11/72\pi^{2}$ and $c=17/720\pi^{2}$, the phase
diagram shows a singular attractor.

Generally, the cosmological constant should be required to realize the de Sitter spacetime.
However, the conformal anomaly provides a new de Sitter attractor and
gives non-trivial effects on the de Sitter spacetime.
Fig.\ref{Fig:phase diagram_cos11} describes the Hubble phase diagram with 
$2 b'\Lambda/3M_{\mathrm{P}}^2 =10^{-0.7}$ for $b'=8/45\pi^{2}$, $c=-1/36\pi^{2}$
and $b'=11/72\pi^{2}$, $c=17/720\pi^{2}$. 
On the other hand,
Fig.\ref{Fig:phase diagram_cos12} shows the Hubble phase diagram with 
$2 b'\Lambda/3M_{\mathrm{P}}^2 =10^{-2.0}$. 
Note that the de Sitter solution of $c >0$ is stable 
under small perturbations of the Hubble parameter with $H\approx \Lambda^{1/2}$
whereas the de Sitter solution of $c<0$ is unstable for $H\ll \Lambda^{1/2}$.
Strangely, the existence of non-zero cosmological constant for $c<0$
does not stabilize (quasi) de Sitter spacetime. 
This result suggests that 
the continuousness of inflation depends on initial conditions for the Hubble paramter.
The stability of classical de Sitter solution can be determined by the sign of 
coefficient $c$,
\begin{align}
c>0\, \Longrightarrow \, N_{G} > \frac{1}{18}N_{S}+ \frac{1}{3}N_{F}.
\end{align}

\begin{figure*}[t]
        \begin{tabular}{cc}
	\begin{minipage}{0.5\hsize}
		\centering
		\subfigure[$b'=8/45\pi^{2}$ and $c=-1/36\pi^{2}$]{
		\includegraphics[width=75mm]{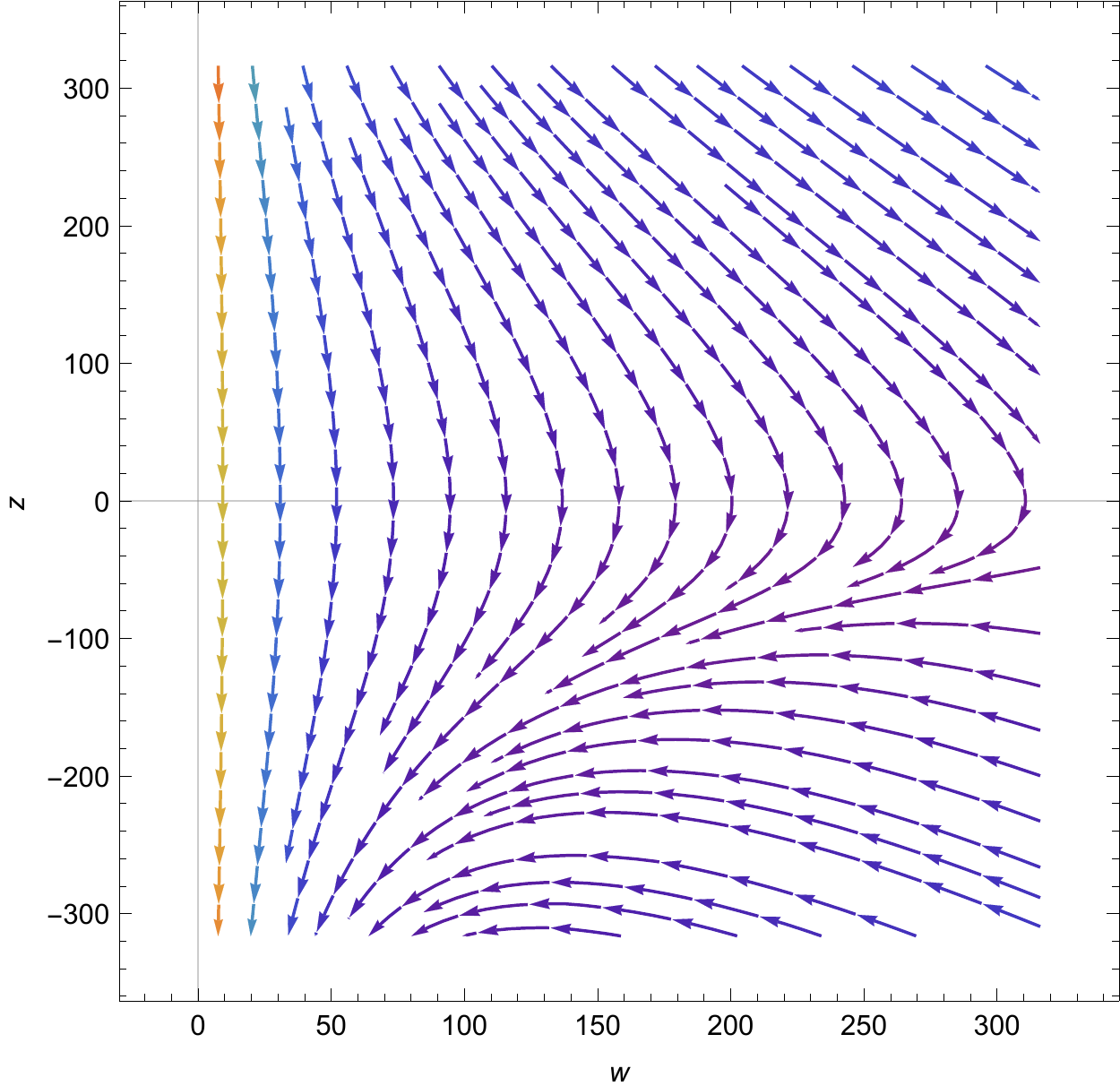}
		\label{fig:MSSMl}}\end{minipage}
\begin{minipage}{0.5\hsize}
		\centering
		\subfigure[$b'=11/72\pi^{2}$ and $c=17/720\pi^{2}$]{
		\includegraphics[width=75mm]{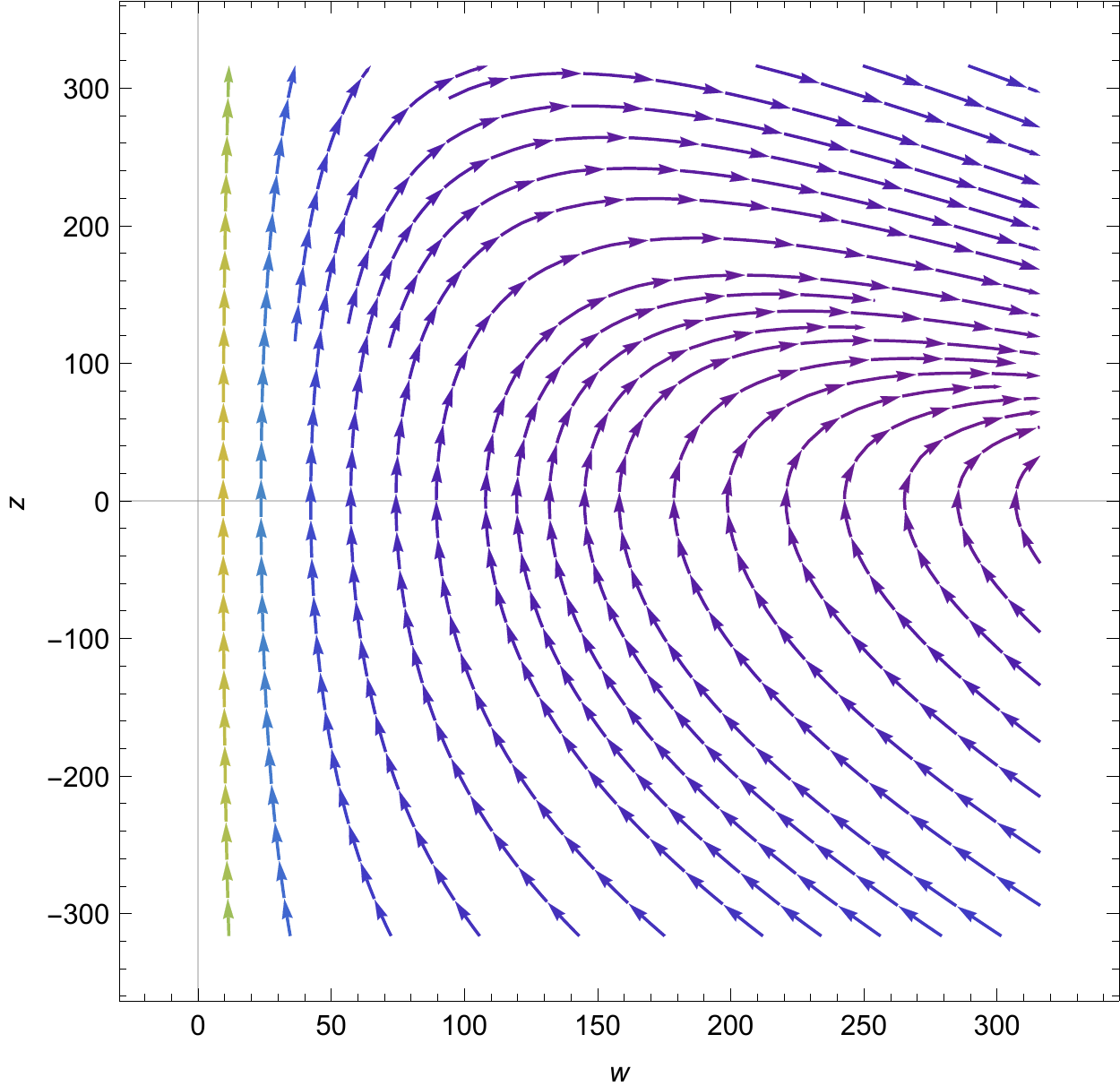}
		\label{fig:SMl}}
\end{minipage}
\end{tabular}
\caption{Phase diagram on Eq.~(\ref{eq:phased_approx}) as a function of $w$ and $z$.  
  Left figure shows the Hubble phase diagram with the cosmological constant
  $\Lambda/M_{\mathrm{P}}^2 =10^{-5.0}$
  and assume $b'=8/45\pi^{2}$, $c=-1/36\pi^{2}$. 
  Right figure shows the Hubble phase diagram with the cosmological constant
  $\Lambda/M_{\mathrm{P}}^2 =10^{-5.0}$
  and set $b'=11/72\pi^{2}$, $c=17/720\pi^{2}$. }
\label{Fig:phase diagram_approx}
\end{figure*}

Next, let us discuss the dynamics of spacetime for $H \approx \Lambda^{1/2}$ and
consider the case $\Lambda^{1/2} \ll  M_{\mathrm{P}}$ in more detail.
For convenience, we rewrite Eq.~(\ref{eq:phased_cos}) using new variables $w$ and $z$:
\begin{align}
\label{eq:phased_cos2}
\frac{dw}{d \tau'}=3 w^{2/3}z, \quad
\frac{dz}{d \tau'}=
\frac{b'\, (w^{5/3}-\frac{M^2_{\rm P}}{2b'\Lambda}w^{1/3} +\frac{M^2_{\rm P}}{6b'\Lambda}w^{-1})}{2c} - 3\, w^{2/3}z,
\end{align}
where:
\begin{align}
\nonumber
w= \Big(\frac{H}{\Lambda^{1/2}}\Big)^{3/2},\quad
z = \frac{\dot{H}}{2 \Lambda^{3/4}} \, H^{-1/2},\end{align}
where we introduce $\tau'=t\cdot \Lambda^{1/2}$.
For simplicity we approximate Eq.~(\ref{eq:phased_cos2}) as follows: 
\begin{align}
\label{eq:phased_approx}
\frac{dw}{d \tau'}=3 w^{2/3}z, \quad
\frac{dz}{d \tau'} \approx
\frac{M^2_{\rm P}}{12\, c\, \Lambda}w^{-1}- 3\, w^{2/3}z,
\end{align}
which clearly show the strong dependence on the sign of 
coefficient $c$.
Fig.\ref{Fig:phase diagram_approx} describes the Hubble flow diagram with 
$\Lambda/M_{\mathrm{P}}^2 =10^{-5.0}$ for $b'=8/45\pi^{2}$, $c=-1/36\pi^{2}$
and $b'=11/72\pi^{2}$, $c=17/720\pi^{2}$. 
The de Sitter solution of $c < 0$ 
breaks down the slow-roll condition of Eq.~(\ref{eq:slow}) eventually
whereas the de Sitter solution of $c > 0$ converges the classical de Sitter attractor
$H_{\mathrm{C}} \simeq \sqrt{{\Lambda}/{3}}$. 
Generally, the quasi de Sitter solution is destabilized by the conformal anomaly
as described by Fig.\ref{Fig:phase diagram_cos11}
and can settle down the classical de Sitter attractor 
only for the specific conditions.
Here, let us summarize conclusions obtained by the above discussion as follows:
\begin{itemize}
\item For $c <0$ and $H(t_0) \lesssim \Lambda$, the de Sitter solutions 
are generally destabilized and the expansion of spacetime terminates: $H(t) \rightarrow 0$.
\item For $c <0$ and $H(t_0) \gtrsim  \Lambda$, the de Sitter solutions 
approach the stable critical point corresponds to the quantum de Sitter attractor: 
$H(t) \rightarrow M_{\mathrm{P}}/\sqrt{2 b'}$.
\item For $c >0$ and $H(t_0) \ll  \Lambda$, the de Sitter solutions go towards the infinity
and the de Sitter expansion of spacetime increases continuously: $H(t) \rightarrow \infty$.
\item For $c >0$ and $\Lambda \lesssim H(t_0) \lesssim M_{\mathrm{P}}/\sqrt{2 b'}$, 
the de Sitter solutions 
approach the stable critical point corresponds to the classic de Sitter attractor: 
$H(t) \rightarrow \sqrt{ {\Lambda}/{3}}$.
\item For $c <0$ and $M_{\mathrm{P}}/\sqrt{2 b'}\lesssim H(t_0)$, 
the de Sitter solutions go towards the infinity
and the de Sitter expansion of spacetime increases continuously: $H(t) \rightarrow \infty$.
\end{itemize}
where we assume $\dot{H}(t_0)=0$ at the initial time for simplicity.
For only $c >0$ and $\Lambda \lesssim H(t_0) \lesssim M_{\mathrm{P}}/\sqrt{2 b'}$, 
the quasi de Sitter solutions can be stable under small metric perturbations
and the spacetime does not change drastically.

\begin{figure*}[t]
        \begin{tabular}{cc}
	\begin{minipage}{0.5\hsize}
		\centering
		\subfigure[$c + 6\left({a_1}-\frac{a_2}{4}\right)=-10^{-2}$]{
		\includegraphics[width=75mm]{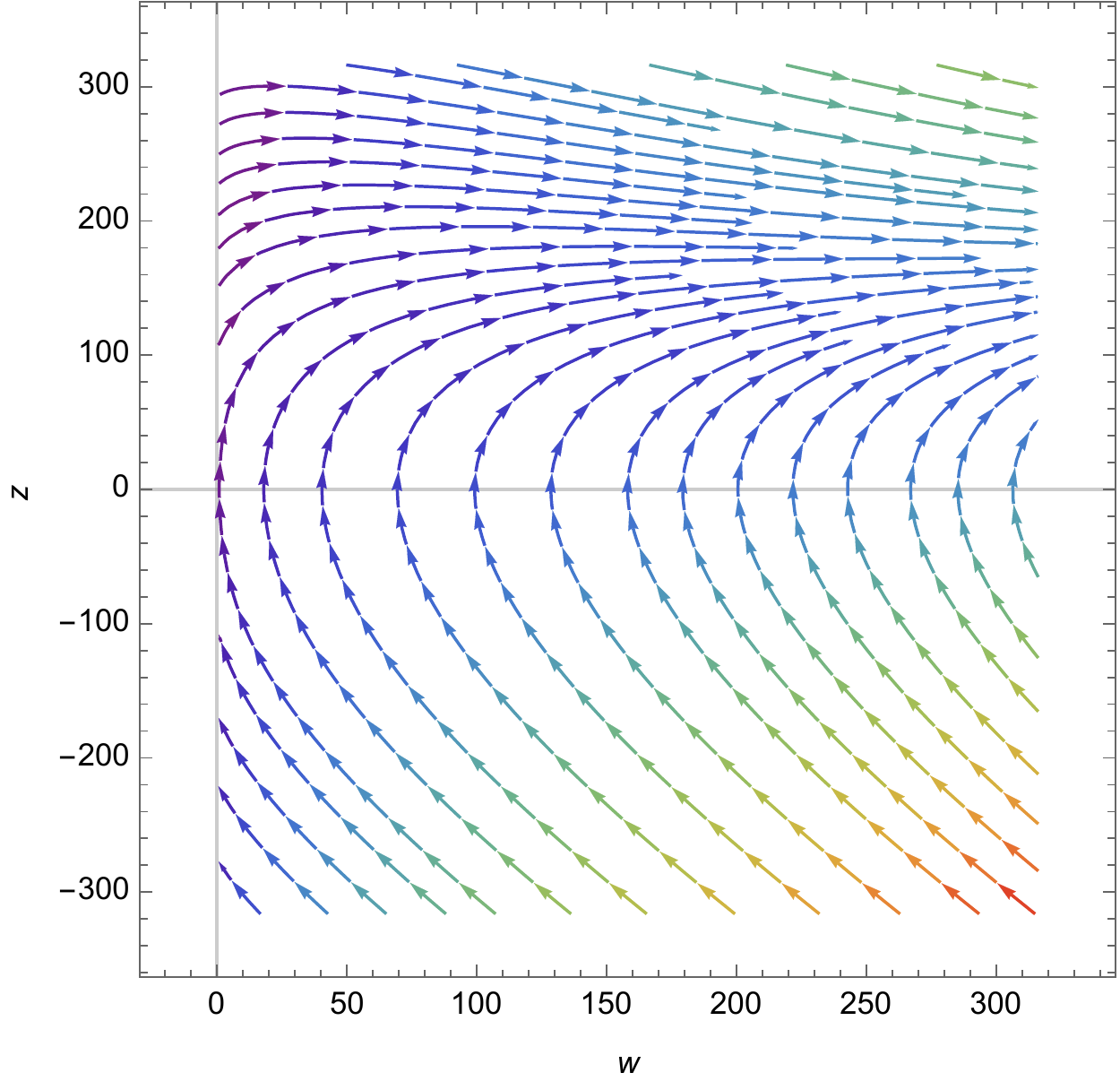}
		\label{fig:MSSMl}}\end{minipage}
\begin{minipage}{0.5\hsize}
		\centering
		\subfigure[$c + 6\left({a_1}-\frac{a_2}{4}\right)=+10^{-2}$]{
		\includegraphics[width=75mm]{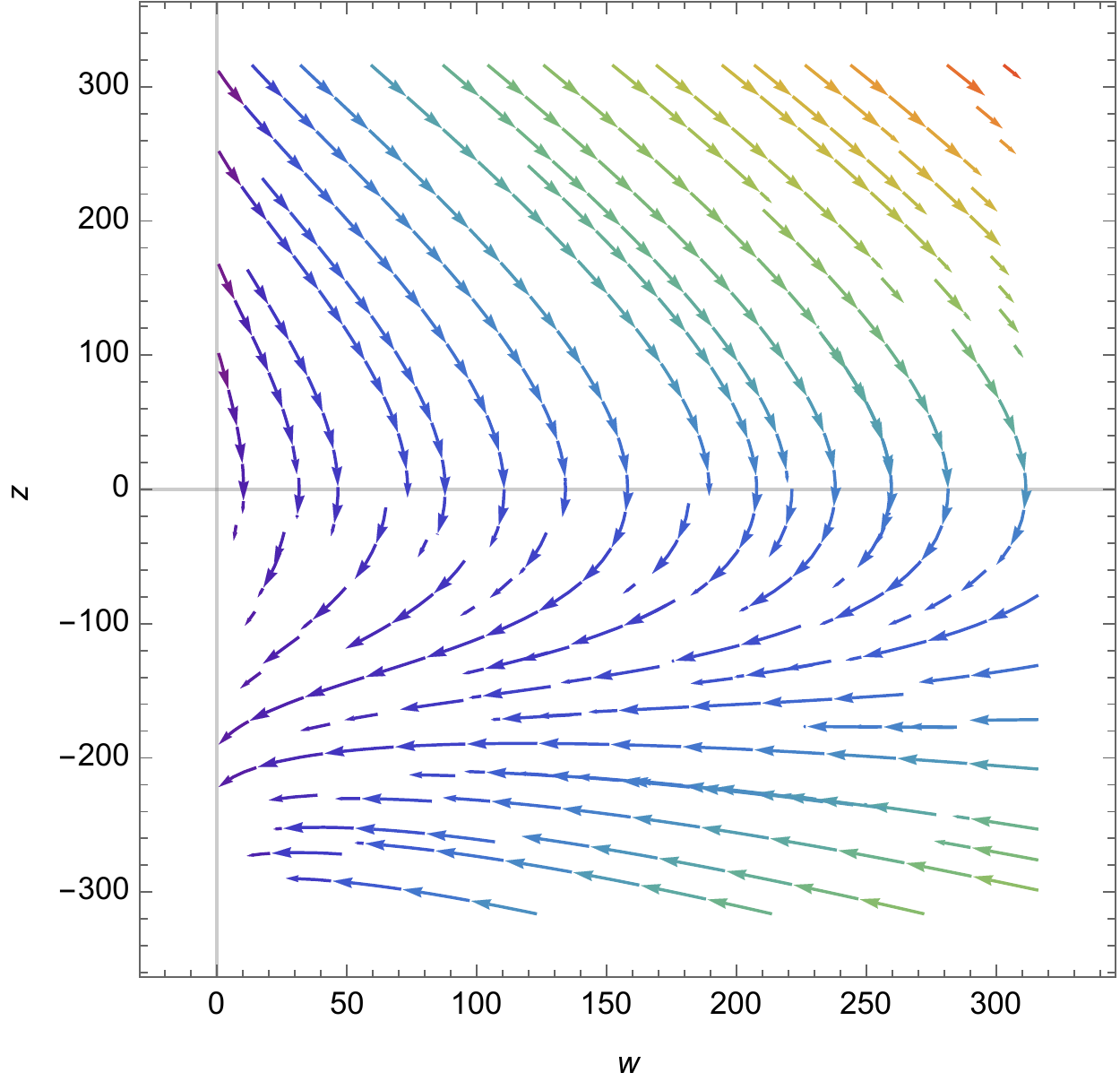}
		\label{fig:SMl}}
\end{minipage}
\end{tabular}
\caption{Phase diagram on Eq.~(\ref{eq:phased_cos3}) as a function of $w$ and $z$
where we set $b' - \left(\frac{a_2}{8}+\frac{a_3}{2}\right)=-10^{-2.0}$ and $\Lambda/M^2_{\rm P}=10^{-2}$.
Left figure shows the Hubble phase diagram with $c + 6\left({a_1}-\frac{a_2}{4}\right)=-10^{-2}$. 
Right figure shows the Hubble phase diagram with $c + 6\left({a_1}-\frac{a_2}{4}\right)=+10^{-2}$. }
\label{Fig:phase diagram_approx2}
\end{figure*}

Finally, let us include the corrections of high-order derivative terms 
${ H }_{ \mu \nu  }^{ \left( 1 \right)  }$, ${ H }_{ \mu \nu  }^{ \left( 2 \right)  }$
or ${ H }_{ \mu \nu  }^{ \left( 3 \right)  }$ for the Hubble flow dynamics.
The trace of effective Einstein's equations of Eq.~(\ref{Einstein's equations}) is given by
\begin{align}
\begin{split}\label{eq:Req}
{ R }-4{ \Lambda  }&={ a }_{ 1 }  { g }^{ \mu \nu  }{ H }_{ \mu \nu  }^{ \left( 1 \right)  }
+{ a }_{ 2 } { g }^{ \mu \nu  }{ H }_{ \mu \nu  }^{ \left( 2 \right)  } + { a }_{ 3 } { g }^{ \mu \nu  }{ H }_{ \mu \nu  }^{ \left( 3 \right)  }
-\left\langle T_{\phantom{\mu}\mu}^{\mu}\right\rangle \\
&= -\left(\frac{a_2}{4}+a_3\right) \left(R_{\mu\nu}R^{\mu\nu}-{1 \over 3}R^2 \right) -6\left(a_1-\frac{1}{4}a_2\right)\Box R
-\left\langle T_{\phantom{\mu}\mu}^{\mu}\right\rangle \\
&= \left(\frac{a_2}{8}+\frac{a_3}{2}\right)E -6\left(a_1-\frac{1}{4}a_2\right)\Box R
-\left\langle T_{\phantom{\mu}\mu}^{\mu}\right\rangle ,
\end{split}
\end{align}
where the conformally anomaly can be removed by the fast and 
second term on the right hand side of Eq.~(\ref{eq:Req}).
Let us rewrite Eq.~(\ref{eq:phased_cos2}) including 
the high-order derivative corrections of ${ H }_{ \mu \nu  }^{ \left( 1 \right)  }$, ${ H }_{ \mu \nu  }^{ \left( 2 \right)  }$
and ${ H }_{ \mu \nu  }^{ \left( 3 \right)  }$ as follows:
\begin{align}
\label{eq:phased_cos3}
\frac{dw}{d \tau'}&=3 w^{2/3}z, \\
\frac{dz}{d \tau'}&=
\frac{\left(b' - \left(\frac{a_2}{8}+\frac{a_3}{2}\right) \right) \left(w^{5/3}-\frac{M^2_{\rm P}}
{2\Lambda\left(b' - \left(\frac{a_2}{8}+\frac{a_3}{2}\right) \right)}w^{1/3} 
+\frac{M^2_{\rm P}}{6\Lambda\left(b' - \left(\frac{a_2}{8}+\frac{a_3}{2}\right) \right)}w^{-1}\right)}
{2\left(c + 6\left({a_1}-\frac{a_2}{4}\right) \right)} - 3\, w^{2/3}z, \nonumber
\end{align}
where we allow that all possibilities of $a_{1,2,3}$ exist
since we do not know the physical value of these parameters.
The high-order derivative terms can provide a negative contribution against $b'$.
In Fig.\ref{Fig:phase diagram_approx2} we show that the Hubble phase diagram 
on Eq.~(\ref{eq:phased_cos3}) as a function of $w$ and $z$.
For simplicity, we set these parameters to be 
$b' - \left(\frac{a_2}{8}+\frac{a_3}{2}\right) =-10^{-2.0}$ and $\Lambda/M^2_{\rm P}=10^{-2}$.
In the left figure we assume $c + 6\left({a_1}-\frac{a_2}{4}\right)=-10^{-2}$
whereas we set $c + 6\left({a_1}-\frac{a_2}{4}\right)=+10^{-2}$ in the right figure.
Note that the de Sitter solutions of $c > 0$
are unstable under small metric perturbations
for $b' - \left(\frac{a_2}{8}+\frac{a_3}{2}\right) < 0 $.
Here, let us summarize the stability condition of the quasi 
de Sitter solution as follows:
\begin{align}
\Lambda \lesssim H(t_0) \lesssim M_{\mathrm{P}} \quad 
\left\{ b' - \left(\frac{a_2}{8}+\frac{a_3}{2}\right) \right\} 
\left\{ c + 6\left({a_1}-\frac{a_2}{4}\right) \right\}  > 0,
\end{align}
which ensures the stability of the classic de Sitter attractor and
the existence of de Sitter spacetime. The quantum gravitational effects involving the conformal anomaly 
and the high-order derivative terms generically destabilize the spacetime and
the cosmic inflation can not last long under reasonable assumptions.

\section{De Sitter instability and eternal inflation }
\label{sec:cosmic_inflation}
In this section we briefly discuss the cosmological application of the de Sitter instability
and especially focus on eternal inflation.
The inflation
assume that our universe experienced a 
quasi de Sitter expansion~\cite{Starobinsky:1979ty,Guth:1980zm,Linde:1981mu,Albrecht:1982wi} and provides elegant solutions 
for the horizon, flatness problems, and 
also generate seeds of primordial density perturbations which 
finally constructs galaxies and large-scale structure.
It is the most reasonable scenarios for the early universe.
However, the inflation is generically eternal~\cite{Guth:2007ng} 
under reasonable assumptions.
The eternal inflation has been classified into three types,
old inflation, new inflation and chaotic inflation models.
Here, we discuss whether the de Sitter instability considered in previous 
Section~\ref{sec:instability} restricts the eternal 
old inflation, new inflation and chaotic inflation.

The old and new inflationary models are 
thought to be generically eternal~\cite{1983veu..conf..251S,Vilenkin:1983xq}.
These models assume that 
some patches of the early universe were the false vacuum
and slowly rolled down the true vacuum state.
The patches of the true vacuum stop the de Sitter expansion and 
the vacuum energy transfers into a hot dense plasma. 
But the entire universe does not stop the inflation.
Although the false vacuum finally collapses, 
the decay of the false vacuum is a suppressed process 
but most of the patches expands exponentially.
In this sense, the inflation of the entire universe
does not come to an end and is usually eternal.
The process creates an infinite number of local universes with the true vacuum 
are often called bubble or pocket universes. 
Thus, it is considered that 
once inflation happens, it creates an infinite number of universes
and leads to the multiverse.
However, the de Sitter instability from quantum backreaction 
could drastically change 
the picture of the multiverse. 
The false vacuum state has the positive cosmological constant 
which cause the de Sitter expansion of the entire universe.
However, the quantum backreaction can destabilize the de Sitter expansion 
and terminate the inflation of the false vacuum state.
Therefore, the old and new inflation could not be eternal 
if the de Sitter spacetime is destabilized from quantum backreaction.

\begin{figure*}[t]
\includegraphics[width=15cm]{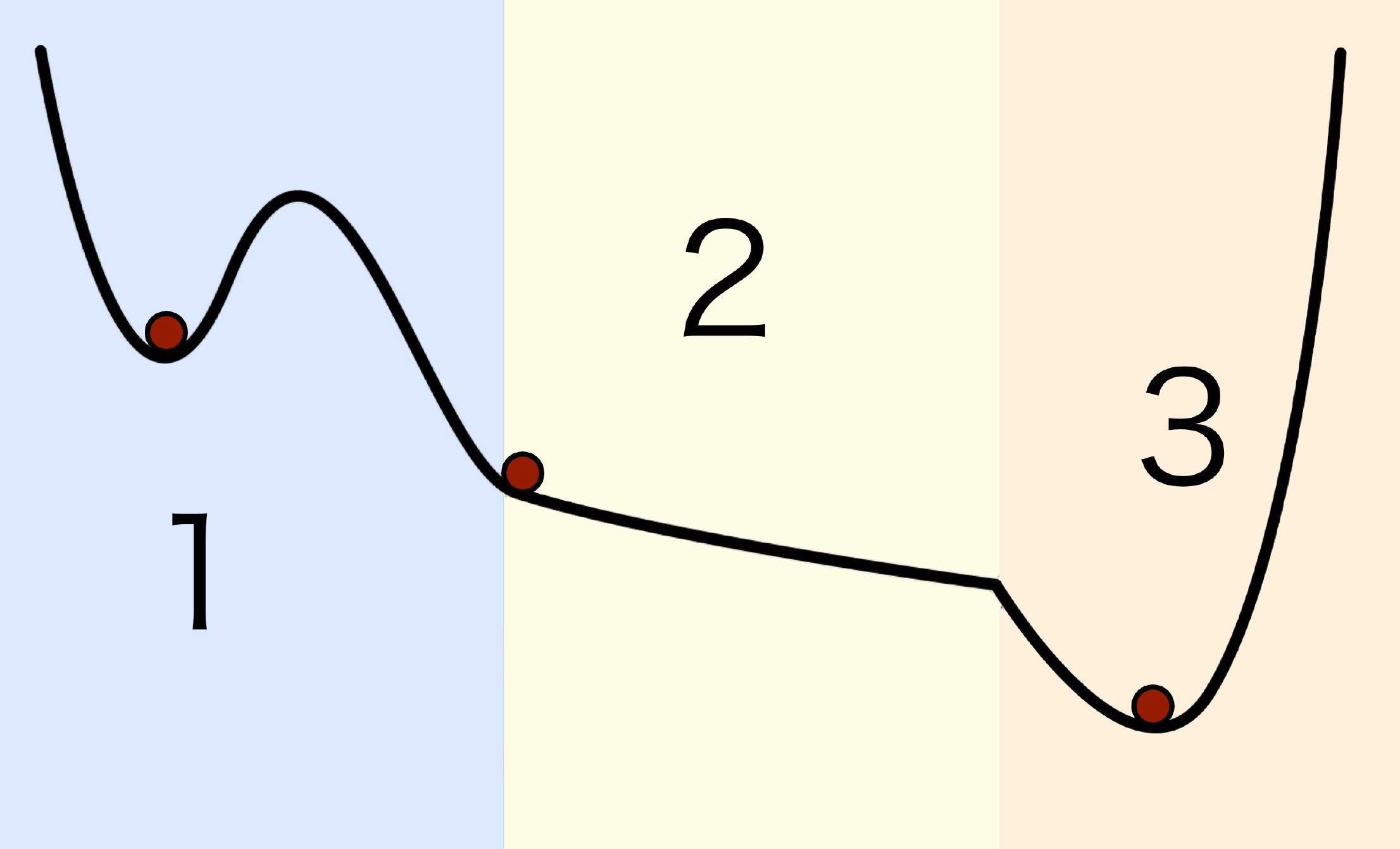}
\caption{A schematic picture of eternal old inflation. 
Region 1 proceeds quantum tunneling from false vacuum into 
low-energy vacuum state. Region 2 proceeds slow-roll inflation.
Region 3 ends inflation and begins reheating.  }
\label{Fig:oldinflation}
\end{figure*}

Let us discuss a relation of the eternal old (new)
inflation and the de Sitter instability in more detail.
We consider the inflaton potential of Fig.\ref{Fig:oldinflation} 
as a example of the old inflation models~\cite{Wainwright:2013lea} 
and review the false vacuum decay in de Sitter spacetime.
The process of the gravitationally induced vacuum decay
can be describe by the Coleman-de Luccia (CdL) formalism~\cite{Coleman:1980aw} 
which corresponds to the  vacuum bubble nucleation.
The vacuum decay rate in curved spacetime is given by~\cite{Coleman:1980aw} 
\begin{align}
\Gamma_{\rm decay } = A\exp(-B),
\end{align}
where $A$ is a prefactor and $B$ is given by the difference between the action of the bounce solution 
and the action of the false vacuum as follows:
\begin{equation}
B =  S_E(\phi) - S_E(\phi_{\rm fv}),
\end{equation}
which is determined by the Euclidean action:
\begin{align}
S_E[\phi,g_{\mu\nu}] = \int d^4x \sqrt{-g}\left[\frac{1}{2}\nabla_{\mu}\phi\nabla^{\mu}\phi + V(\phi) - \frac{M_{\rm{P}}^2}{2}R\right].\label{eq:EuclideanAction}
\end{align}
To discuss instanton mediated vacuum transitions in curved spacetime we consider the
Euclidean analogue of cosmological spacetime: $d s^2 = d\chi^2 + a^2(\chi)d\Omega_3^2$
where $\chi^2 = t^2+r^2$, $a(\chi)$ is the Euclidean scale factor 
and $d\Omega_3^2$ is the metric of a 3-sphere.
The equations of motion in this case are written as~\cite{Rajantie:2017ajw}
\begin{align}
\ddot{\phi} &+ \frac{3\dot{a}}{a}\dot{\phi} - V'(\phi) = 0,  \\
\dot{a}^2 &= 1 + \frac{a^2}{3M_{\rm{P}}^2}\left(\frac{\dot{\phi}^2}{2} - V(\phi)\right).
\end{align}
Solving these equations the decay exponent $B$ can be taken as
\begin{align}
B = \frac{24\pi^2 M_{\rm{P}}^4}{V(\phi_{\rm{fv}})} - 2\pi^2\int_{0}^{ \infty}d\chi a^3(\chi)V(\phi(\chi))
\end{align}
The trivial solution of the Euclidean equations of motion in de-Sitter spacetime
assumes that the field stays on the top of the potential.
This solution is known as the Hawking-Moss instanton~\cite{Hawking:1981fz},
\begin{align}
\phi(\chi) = \phi_{\rm{max}},\ a(\chi) =\frac { \sqrt { 3 } { M }_{ \rm P }  }{ { V\left( { \phi  }_{\rm  max } \right)  }^{ 1/2 } }
\sin {  \left(\frac { { V\left( { \phi  }_{ \rm max } \right)  }^{ 1/2 } }{ \sqrt { 3 } { M }_{ \rm P } }\chi   \right)  } ,
\end{align}
For this solution the decay exponent $B$ is given by
\begin{align}
B = 24\pi^2M_{\rm{P}}^4\left(\frac{1}{V(\phi_{\rm{fv}})} - \frac{1}{V(\phi_{\rm{max}})}\right)
\approx \frac{8\pi^2 V(\phi_{\rm{max}})}{3H^4},
\end{align}
which represents the probability that thermal fluctuation with the Gibbons-Hawking temperature $T_{\rm GH}=H/2\pi$
pushes the inflaton field $\phi$ from the false vacuum to the top of the potential. 
On the other hand the Coleman-de Luccia (CdL) instanton can be interpreted as that thermal fluctuation pushes $\phi$
partially and the pushed $\phi$ goes out to true vacuum through quantum tunneling.
If the potential barrier is compared with the Hubble scale,
the CdL instanton does not necessarily exist~\cite{Batra:2006rz} and the transition is described by the Hawking-Moss instanton.
Note that the Hawking-Moss transition should be interpreted as an entire Hubble-volume tunneling~\cite{Batra:2006rz,HenryTye:2008xu},
and the vacuum transition occurs on only one Hubble patch not the entire universe, and moreover,
it is a exponentially suppressed process.
However, the number of the Hubble patches exponentially increases,
\begin{align}
\mathcal{N}_{\rm patch}\sim \exp\left( 3Ht \right)
\end{align}
which overcomes the vacuum decay rate in de Sitter spacetime.
Thus, the number of the false vacuum patches exponentially increases
with Hubble time, 
\begin{align}
\mathcal{N}_{\rm inflation }\sim \mathcal{N}_{\rm patch} 
\cdot (1-\Gamma_{\rm decay })^{Ht}
\sim e^{Ht \cdot \left\{3 + \ln {(1- \Gamma_{\rm decay })} \right\}} 
\gg \mathcal{O}(1),
\end{align}
and the inflation as a whole universe is always eternal.
However, the quantum gravitational effects involving the conformal anomaly
can destabilize the de Sitter expansion of the background spacetime
as previously discussed in Section~\ref{sec:instability}. For instance 
in the case $H(t) \rightarrow 0$ the inflation of the false vacuum terminates 
and therefore the eternal new inflation does not happen.
On the other hand in the case $H(t) \rightarrow M_{\mathrm{P}}$ 
the gravitational false vacuum decay is enhanced
and the backreaction of the thermal fluctuation of the cosmological horizon with $T_{\rm GH}=H/2\pi$
to the inflaton potential can not be ignored as follows:
\begin{align}
V\left( \phi  \right) \longrightarrow V\left( \phi  \right) +\mathcal{O}\left(T_{\rm GH}^2\right)\phi^2
+\mathcal{O}\left(T_{\rm GH}^4\right).
\end{align}
where $T_{\rm GH}\rightarrow M_{\mathrm{P}}$.
The de Sitter thermalization or quantum backreaction could modify the 
inflaton potential or lead to the spontaneous symmetry restoration~\cite{Hawking:1980ng}
on the entire Universe.
Thus, the quantum gravitational effects drastically changes 
scenarios of the old/new inflation and they are not eternal 
without the specific conditions.

Next let us discuss eternal chaotic inflation models.
The chaotic inflation can also be eternal 
by quantum fluctuation~\cite{Linde:1986fc,Linde:1986fd,Goncharov:1987ir}.
The quantum fluctuation 
on the de Sitter spacetime can be considered as a brownian fluctuation 
of $H/2\pi$ at time intervals $H^{-1}$.
The each Hubble patches experience different random walks and 
the vacuum expectation value $\left< { \delta\phi  }^{ 2 } \right>$ varies in each patches.
\begin{align}
\left< { \delta\phi  }^{ 2 } \right> =\frac{ { H }^{ 3 }t }{ 4{ \pi  }^{ 2 } }.
\end{align}
We assume regions separated by a distance $l$ and definite the distance  time $t_{l}$
as $l=H^{-1}\exp H\left( t-t_{l} \right)$. When the distance $l$ becomes larger than horizon-size, 
the field values $\phi$ becomes different by the quantum fluctuation.
The mean-square field variation for the distance $l$ is written as
\begin{align}
\left< { \delta\phi  }^{ 2 } \right> \sim \frac{ { H }^{ 3 } }{ 4{ \pi  }^{ 2 } }\left( t-{ t }_{ l } \right) \sim \frac{ H^{2} }{ 4\pi^{2}  }\log { \left( Hl \right)  } 
\end{align}
Due to this variation, the reheating process does not occur simultaneously in different region of the entire universe.
The presently observable universe is sufficiently homogeneous and isotropic.
Therefore, the region of the visible universe should have been simultaneously thermalized 
with very small fluctuation although the universe on much large scale have large deviations 
from homogeneity and isotropy~\cite{Aryal:1987vn}. 
We consider the probability $P\left(\phi,t\right)$ in one Hubble patch $\phi<H$ at time t.
The scalar filed $\phi$ exhibits brownian motion of step $+H/2\pi$ and $-H/2\pi$. 
The probability $P\left(\phi,t\right)$ in one Hubble patch is given as follows,
\begin{align}
P\left(\phi,t\right)\sim \left(\frac{1}{2}\right)^{Ht}=e^{-Ht\ln {2} }.
\end{align}
Thus, the probability $P\left(\phi,t\right)$ in one Hubble patch 
is very small at the sufficient time. 
Most of the Hubble patches ends the inflation and begins the reheating. 
But, the total number of Hubble patches increases $\mathcal{N}_{\rm patch}\sim \exp\left( 3Ht \right)$ and the number of Hubble patches continuing the inflation 
grows exponentially with time,
\begin{align}
\mathcal{N}_{\rm inflation}\sim \mathcal{N}_{\rm patch}
\cdot P\left(\phi,t\right) =e^{Ht\cdot\left(3 - \ln {2} \right)}\sim e^{2.3Ht}.
\end{align}
Most of the Hubble patches of the universe 
are still during the inflation and the whole universe is eternally expanding. 
Our universe is regarded as a small region of the whole universe 
which has deviated from the eternal inflation. 
However, the de Sitter instability from quantum backreaction involving the conformal anomaly 
restrict the possibility of the eternal chaotic inflation.
For instance 
in the case: $H(t) \rightarrow 0$ the inflation of these patches terminates 
and therefore the eternal inflation would not happen.
On the other hand in the case: $H(t) \rightarrow M_{\mathrm{P}}$, 
the inflaton potential acquires the effective mass $\mathcal{O}\left(T_{\rm GH}^2\right)\phi^2$
and might break down the slow-roll condition although 
the scenario in the case of chaotic inflation strongly depends on the potential.

Let us summarize the argument in Section~\ref{sec:cosmic_inflation}.
The quantum gravitational effects involving the conformal anomaly and the high-order derivative corrections
destabilize the de Sitter spacetime and 
drastically change a eternal picture of old, new and chaotic inflation models.
For $H(t) \rightarrow 0$, the inflation inevitability terminates.
For $H(t) \rightarrow M_{\mathrm{P}}$,
the inflation strongly depends on the potential.

\section{Conclusion and Summary}
\label{sec:conclusion}
In the present paper we have discussed the de Sitter instability 
from quantum backreaction involving the conformal anomaly.
The instability of (quasi) de Sitter spacetime from 
quantum gravitational effects has been discussed in many works.
Especially, the gravitational particle production or thermal feature of the
de Sitter spacetime suggest that the (quasi) de Sitter spacetime 
might not be stable.
In order to investigate the stability of the de Sitter spacetime 
we have focused on the conformal field theory (CFT) and discussed 
quantum backreaction involving the conformal anomaly.

The conformal or trace anomaly corresponds to the 
quantum gravitational contributions of the massless conformal fields
and affects the background spacetime homogeneously.
First, we have derived the conformal anomaly using the adiabatic (WKB) approximation
and discussed the renormalization of the quantum energy momentum tensor.
We have clearly shown that the ambiguity of the conformal anomaly
can be reduced by taking the adiabatic (WKB) approximation method.
Then, we have considered the dynamics of the Hubble parameter 
based on the semiclassical Einstein's equations
with the cosmological constant, the conformal anomaly and the higher-derivative terms.
We have clearly shown that the quasi de Sitter solutions are generally unstable 
from the viewpoint of the semiclassical gravity
and it can settle down the classical de Sitter attractor $H_{\mathrm{C}} \simeq \sqrt{{\Lambda}/{3}}$
only for the specific conditions.
We have obtained the stability condition of the classical de Sitter solutions,
\begin{align*}
\Lambda \lesssim H(t_0) \lesssim M_{\mathrm{P}} \quad 
\left\{ b' - \left(\frac{a_2}{8}+\frac{a_3}{2}\right) \right\} 
\left\{ c + 6\left({a_1}-\frac{a_2}{4}\right) \right\}  > 0.
\end{align*}

Our results suggest that the quasi de Sitter spacetime is not stable 
and the inflation is destabilized except for the specific conditions. 
Under reasonable assumptions 
the inflation finally becomes the Planckian inflation with the Hubble scale 
$H \approx M_{\rm P}\equiv \sqrt{1/8\pi G_{N}}$ or terminates $H(t) \rightarrow 0$. 
Therefore, the quantum gravitational effects with the conformal anomaly and the high-order derivative corrections
drastically change a picture of the eternal old, new and chaotic inflation.
For $H(t) \rightarrow 0$ the inflation inevitability terminates.
For $H(t) \rightarrow M_{\mathrm{P}}/\sqrt{2 b'}$ or $H(t) \rightarrow \infty$
the final state of the inflation strongly depends on the potential.

\section*{Acknowledgments}
I would like to thank Fuminobu Takahashi for discussions and longterm collaborations 
on this research, and I am also grateful to Masahiro Hotta, Atsushi Naruko and 
Naoki Watamura for giving valuable comments and discussions.

\appendix
\section{Geometrical tensors in FLRW metric}
In the FLRW metric, 
the Ricci tensor and the Ricci scalar are given as follows:
\begin{align}
\begin{split}
R_{00}&=-\frac { 3 }{ 2 }D',\quad R_{11}=\frac { 1 }{ 2 }\left(D'+D^{2}\right),
\quad R=\frac{3}{C}\left({ D' +\frac { 1 }{ 2 }  { D }^{ 2 } } \right),\\
G_{00}&=\frac { 3 }{ 4 }D^{2},\quad G_{ii}=-D'-\frac { 1}{ 4 }D^{2}, \\
H_{00}^{\left(1\right)}&=-\frac { 9 }{ C }\left( \frac { 1 }{ 2 }D'^{2}-D''D +\frac{3}{8}D^{4} \right), \\
H_{ii}^{\left(1\right)}&=-\frac { 3 }{ C }\left( 2D'''-D''D+\frac { 1 }{ 2 }D'^{2}-3D'D^{2} +\frac{3}{8}D^{4} \right).
\end{split}
\end{align}

\bibliographystyle{JHEP}
\bibliography{dsinstability}
\end{document}